\documentclass[12pt]{article}
\usepackage{amssymb}
\usepackage{amstext}
\newcommand*\de{\mathrm{d}}
\newcommand*\De{\mathrm{D}}
\renewcommand*\epsilon{\varepsilon}
\renewcommand*\phi{\varphi}
\renewcommand*\theta{\vartheta}
\begin{document}

\title{  The Higgs sector of gravitational gauge theories} 
 
\author{M. Leclerc \\ {\small \it Section of Astrophysics and Astronomy, 
Department of Physics,} \\ \small \it University of Athens, Greece}  
\date{}
\maketitle
\textbf{Abstract} \\[2mm]
{\footnotesize Gravitational gauge theories with  de Sitter, 
Poincar\'e and affine symmetry group   
are investigated under the aspect of the breakdown of the initial symmetry 
group down to the Lorentz subgroup. As opposed to the nonlinear 
realization approach, in the dynamical symmetry breaking procedure, 
the structure subgroup  
is not chosen arbitrarily, 
but is dictated by the symmetry of the groundstate of a Higgs field. 
We review the theory of spontaneously broken de Sitter gravity 
by Stelle and West and apply a similar approach to the case of 
the Poincar\'e and affine groups. 
We will find that the Poincar\'e case is almost trivial. The  
translational 
Higgs field  
reveals itself as pure gauge, i.e., it is expressed entirely 
in terms of the Nambu-Goldstone bosons and does not appear in the  Lagrangian
after the symmetry breaking. 
The same holds for the translational part of the affine group. 
The Higgs field provoking the breakdown of the general linear group 
leads to the determination of the Lorentzian signature of the metric 
in the groundstate. We show that the Higgs field  remains in its 
groundstate, i.e., that the metric will have Lorentzian signature, unless 
we introduce matter fields that explicitely couple to the symmetric part of  
the connection. Furthermore, we present arguments that the Lorentzian 
signature is actually the only possible choice for physical spacetime, 
since the symmetry breaking mechanism works only if the stability 
subgroup is taken to be the Lorentz group. The other four-dimensional 
rotation groups are therefore  ruled out not only on physical, but 
also on theoretical grounds. Finally, we show that some features, like 
the necessity of the introduction of a dilaton field, 
that seem artificial in the context of the affine theory, appear most  
natural if the gauge group is taken to be the special linear group in 
five dimensions. We also present an alternative model which is based 
on the spinor representation of the Lorentz group and is especially 
adopted to the description of spinor fields in a general linear 
covariant way, without the use of the infinite dimensional representations   
which  are usually considered to be unavoidable.}


\section{Introduction}

That classical gravity represents 
the subsidiary of a theory with larger symmetry has been suggested 
in the past, presenting the theory in terms of nonlinear realizations 
of those groups. 
This procedure, however, requires that 
the Lorentz subgroup is chosen arbitrarily as structure group 
and therefore, does not explain the Minkowskian structure  of the physical 
spacetime. On the other hand, introducing explicitly a Higgs sector 
into the theory, results in a dynamical symmetry breaking, with a 
groundstate invariant under the Lorentz subgroup only. Such a model 
has been presented by Stelle and West in the case of the de Sitter 
group. We will review their results and apply a similar approach to 
the Poincar\'e and affine groups. 

The concept of nonlinear realizations has been introduced 
shortly after the advent of spontaneously broken theories \cite{1,2,3}
and its most successful candidate, the electroweak theory of Weinberg 
and Salam \cite{4}, by Coleman, Wess and Zumino in the context of gauge 
theories of internal symmetry groups  \cite{5} (see also \cite{6}). 
Applying these concepts to the general linear group $GL(R^4)$, Isham, Salam 
and Strathdee \cite{7} formulated the first 
version of a \textit{broken} gauge 
theory of gravity. Since then, many attempts have been made to use  
nonlinear realizations in the context of gravity, see for instance 
\cite{8,9,10} for the (super)Poincar\'e group, \cite{11} for the de Sitter 
group, \cite{12} for the affine group and \cite{13,14} for 
attempts to include diffeomorphism 
groups. The conformal group has also received considerable 
attention \cite{7,13,15,16}. Other references can be found in the cited 
articles. 

In the nonlinear realization approach, the structure (or stability) 
subgroup $H$  
of the symmetry group $G$
can be chosen arbitrarily. This is a result of the fact that the 
choice of a certain subgroup essentially consists in parameterizing some 
of the gauge degrees of freedom. Simply stated, some parts of the 
connection are transformed into tensors with the help of Nambu-Goldstone 
fields 
and the corresponding gauge 
freedom is transferred to those fields. 
This is not a gauge choice or gauge fixing, but rather a 
gauge parameterization. As such, it does not effect the $G$ symmetry of the 
theory in any way, and therefore the choice of one or another subgroup 
is necessarily equivalent. 

Clearly, this is somehow unsatisfactory, since we know from everyday physics
that the residual symmetry group should be the Lorentz group $SO(3,1)$ 
(or eventually the Poincar\'e group) and not, say, 
the de Sitter group or the group $SO(4)$. After all, there is little doubt 
that we live in a 4-dimensional Minkowskian world and not in a 5-dimensional 
or Euclidian one. It is therfore natural to look for a mechanism that 
fixes the choice of the structure group as a result of the dynamics of the 
theory. The introduction of a Higgs field, carrying a representation of 
the symmetry group $G$, with a suitable Higgs potential 
leading to a groundstate that is not symmetric under $G$ anymore, but only 
under the Lorentz subgroup $H=SO(3,1)$ provides such a mechanism.

That gravity is a dynamically broken gauge theory  
has been suggested in the past, but with the exception of the 
work of Stelle and West \cite{11}, we have never really seen neither a 
kinetic term for the Higgs field, nor a potential provoking the 
symmetry breakdown. We refer here only to theories that are 
constructed, as far as possible, without the help of additional 
ingredients, merely in terms of  
the connection of the symmetry group and those fields that are 
unavoidable to write down a meaningful Lagrangian density, as are 
the Poincar\'e coordinates or the metric in the case of the general 
linear group. 

It is most astonishing, especially in the case 
of Poincar\'e gauge theory, which has raised the interest of so many people, 
that although  the introduction of the Nambu-Goldstone field 
(also called Poincar\'e  coordinates, or Cartan's radius vector)
in the manner of  \cite{10,17} and \cite{18}  
(whose ultimate interpretation is provided by 
the nonlinear realization approach, see \cite{10,19}) is widely accepted, 
nobody seems to have presented the Higgs sector that is supposed 
to break down the symmetry from the Poincar\'e to the Lorentz group. 
Let us cite a characteristic statement from the standard 
reference on metric affine gauge theory by Hehl et al. \cite{20}: 
 \textit{We believe that the story of the $\xi$} (the 
Poincar\'e coordinates) \textit{has not yet come to an end and that future 
developments on this point are possible. Probably, one has to come 
up  with an idea of how to construct an explicit symmetry breaking 
mechanism.} We will show that, depending on the degree of optimism 
in the physicist's viewpoint, the Higgs sector for this breaking mechanism is 
actually already implicitly incorporated in the standard Poincar\'e 
gauge theory framework, or, taking a less optimistic viewpoint, 
that no Higgs sector (other than that) can be constructed. 

On the other hand, symmetry breaking mechanisms have been presented 
in the case of the general linear group. However, the Higgs 
sector in those cases is constructed with the help of an additional field in 
 a general linear, infinite dimensional 
representation (a so called manifield) \cite{12,21}. 
This is unsatisfying from several points of view. First, it does 
not explain the role of the general linear metric tensor, which 
is a necessary ingredient of the theory since no Lagrangian 
invariant under the general linear group can be written down without the 
help of a metric. The nonlinear realization approach clearly suggests 
that it is this tensor that should play the role of the Higgs field 
and trigger the symmetry breaking down to the Lorentz group. 
This avoids not only the use of infinite dimensional representations, 
but also the introduction of yet another field to the already 
complicated structure of the theory. 
Secondly, it is not really clear to us how the manifield approach  
 favors the Lorentz group as stability subgroup 
as opposed to  the  
 other possible rotation groups, like $O(4)$ etc.  We will present 
a much simpler approach, with the metric playing the role of the Higgs 
field, and the symmetry breaking fixing the signature of the groundstate 
metric. 

However, it turns out, that this method actually works only in  
the case where the symmetry group is taken to be the special linear 
group. In order to reduce the general linear group to the 
special linear group, we will have to introduce an additional Higgs 
dilaton field, following a similar approach as in \cite{20}. 
We present arguments that explain why the introduction of this 
field (which is not really apparent in the nonlinear realization 
scheme) is unavoidable. In brief, the reason can be traced back 
to   the fact that, 
in the  metric affine theory, one of the field equations is redundant. 
Moreover, we will show that the dilaton approach contains a small 
loophole, and in order to construct a consistent Higgs sector, 
an additional scalar field has to be included. 

It turns out that the introduction of both the scalar and the dilaton 
field can be naturally explained in the context of the gauge 
theory of the special linear group $SL(R^5)$. This theory combines 
the features of the affine and de Sitter theories and provides us 
with a mechanism to break down the symmetry to the Lorentz group 
without any additional fields.  

In order to avoid misunderstandings, it is important to point out 
that throughout the article, the gauge group is treated exactly in 
the same way as in conventional gauge theories of internal 
symmetries. Thus, the gauge potential is a connection one-form 
transforming under the group $G$ in the usual way (see equ. (4) below), 
this gauge group being in no way related to spacetime 
coordinate transformations. The difference between our theories 
and conventional Yang-Mills theories will arise through additional  
structure, like tetrad fields or metrics, that will be defined with 
the help of the connection and of the Higgs fields, a posteriori.
 This conception 
of gauge gravity coincides essentially with the 
one described in \cite{19} and \cite{20}. It seems to us the most promising 
way in view of a possible unification of gravity with the other gauge 
interactions. It is, however, not the only way. Alternative theories, 
based on a different gauge concept, have been presented and analyzed 
in detail in the past. In those theories, the general linear gauge 
group is directly related to the coordinate transformations and the 
tetrad fields arise as the Goldstone fields of the symmetry breakdown.  
In other words, the quotient space $G/H$, where $H$ is the Lorentz group, 
is identified with physical spacetime, and not, as in our approach, 
with a tangent space to the spacetime manifold. We will not deal 
with such theories in this article and refer the reader to 
the original article of Ivanenko and Sardanashvily \cite{22} as well 
as to  more recent work, e.g., \cite{23,24}. 
The later,  
dealing with the coupling of gravity to spinor matter, is     
essentially interesting in connection  with section 7 of this article. 
Note also that 
 the approach of \cite{12,13,14}, and also of one of the earliest Higgs  
approaches to gravity  by Borisov and Ogievetsky \cite{25},  
are  based on a similar gauge  concept 
concerning the general linear group. Without going into details, it is clear 
what are the weak and strong points of both, alternative, viewpoints. 
As outlined above, the approach of Hehl et al., which we adopt here, 
is more suitable in view of a unification of the fundamental forces, 
since, in a sense, it is based on the idea to describe gravity in a way 
as close as possible to the description of the other forces of 
nature. On the other hand, 
the diffeomorphism approach of \cite{22} essentially starts by 
underlining not the similarities to other interactions, but rather the 
 features that are unique to gravity, namely the equivalence principle and the 
universality, which are then married with a suitable gauge concept. 
 This is certainly a promising way too, since from those  same  
principles, Einstein was led  successfully to general relativity in the 
first place.  In our approach,  the validity of those 
principles has to be established a posteriori, while the fundamental 
concept is the gauge principle. This is not necessarily a drawback, since 
it it not evident whether the equivalence principle is valid in generalized 
gravity theories,  or whether it represents just a sort of a classical, 
macroscopic limit. For instance, in Poincar\'e gauge theory, 
as is well known, the equivalence principle is certainly not 
generally valid, 
since particles with different intrinsic spin follow different 
trajectories \cite{26}. Interesting remarks concerning 
 the different gauge concepts  are also found in \cite{27}.     

We will begin this article by reviewing the gauge theory of the 
de Sitter group  
as it was presented by Stelle and West twenty five years ago. This 
gives us the possibility to follow step by step the same method in the 
other cases, which present some particularities, and which 
 else would hardly  be recognizable  as dynamically broken theories. 

In detail, the article is organized as follows. 
After a short excursion to the nonlinear realization approach (section 2), 
we will take the work of Stelle and West as a starting point for the 
construction of the Higgs sector of gauge theories of gravity for 
different symmetry groups. We start by reviewing the de Sitter case 
(section 3) and go on to the Poincar\'e and affine groups in 
sections 4 and 5. Finally, a short analysis of the $SL(R^5)$ theory 
will be given in section 6, focusing on the similarities to the 
affine case. In a last section, we briefly describe an 
alternative Higgs mechanism where the general linear symmetry 
is broken down to the group $SL(2,C)$, i.e., to the spinor representation 
of the Lorentz group.

\section{Nonlinear realizations}

We  very briefly review the basic concepts of this approach, in order to 
compare later on with the conventional Higgs mechanism. A modern and 
compact description of the method can be found in \cite{19}, in the context 
of the affine group.  
 
Let $G$ be the symmetry group of the theory and $H$ the structure subgroup 
(which is chosen arbitrarily in this approach). Consider a section  
 on the principal fiber bundle $G(G/H,H)$, 
\begin{equation}
\sigma: G/H \rightarrow G. 
\end{equation}
Then, any element $g$ of $G$ can be written uniquely as 
\begin{equation}
g = \sigma(\xi) h, 
\end{equation}
with $\xi \in G/H$ and $h\in H$. Especially, the element $g\sigma(\xi)$ can 
be written as 
\begin{equation}
g\sigma(\xi) = \sigma(\xi') h, 
\end{equation}
for some $h\in H$. Obviously, the coset parameters $\xi'$ as well as  
$h$ will depend on $\xi$ and $g$, i.e., $\xi' = \xi'(\xi, g)$ and 
$h = h(\xi,g)$. A short form of (3) is $\sigma' = g \sigma h^{-1}$. 

If we have a linear connection transforming under $G$ as 
\begin{equation}
\Gamma \rightarrow  g\Gamma g^{-1} + g \de g^{-1}, 
\end{equation}
then the following quantity
\begin{equation}
\tilde \Gamma = \sigma^{-1} \Gamma \sigma + \sigma^{-1} \de \sigma, 
\end{equation}
clearly transforms as 
\begin{equation}
\tilde \Gamma \rightarrow h \tilde\Gamma h^{-1} + h \de h^{-1}, 
\end{equation}
i.e., basically as a (nonlinear) connection under the group $H$ (apart from 
those components that are outside of the Lie algebra of $H$, which 
transform as tensors under $H$, see \cite{19}). 

More generally, to any field $\phi$ transforming under some representation of 
G as $\phi \rightarrow g\phi$, we can associate a field $\psi =
\sigma^{-1}\phi$ that will transform as $\psi \rightarrow h \psi$. 
 This presupposes of course that the transformation law $\sigma \rightarrow 
g\sigma h^{-1}$ is interpreted as transformation under the corresponding 
 representations. We will come back to this important point in section 7.

Let us write the abstract formulas (3) to (6) in component form.
Quantities transforming under the group $G$ will be labeled
by greek indices $\alpha, \beta\dots $ and those transforming under 
the subgroup $H$ by latin letters from the beginning of the alphabet 
$a,b,c, \dots $. (We reserve latin letters from the middle of the 
alphabet $i,j,k\dots $ for spacetime indices.) Equation (3) in the form 
$\sigma' = g\sigma h^{-1}$ shows that $\sigma $ transforms as a mixed 
tensor. We will denote it by $r^{\alpha}_{\ a}$ and its inverse by 
$r^a_{\ \alpha}$. Let $\Gamma^{\alpha}_{\ \beta}$ be the $G$-connection
 1-form, 
i.e., we have the transformation law   
\begin{equation}
\Gamma^{\alpha}_{\ \beta} \rightarrow G^{\alpha}_{\ \gamma}
\Gamma^{\gamma}_{\ \delta} (G^{-1})^{\delta}_{\  \beta} + G^{\alpha}_{\ \gamma}
\de (G^{-1})^{\gamma}_{\ \beta}, 
\end{equation}
where $G^{\alpha}_{\ \beta}$ is an element of $G$. 
Equation (5) defines the $H$-connection as
\begin{equation}  
\Gamma^a_{\ b} = r^a_{\ \alpha}
\Gamma^{\alpha}_{\ \beta} r^{\beta}_{\  b} + r^{a}_{\ \alpha}
\de r^{\alpha}_{\ b}. 
\end{equation}
Using the transformation law for $r^{\alpha}_{\ a}$ 
\begin{equation}
r^{\alpha}_{\ a} \rightarrow 
G^{\alpha}_{\ \beta}r^{\beta}_{\ b}(H^{-1})^{b}_{\ a},
\end{equation}
with $H^a_{\ b} \in H$, we find 
\begin{equation}
\Gamma^a_{\ b} \rightarrow \Gamma^a_{\ b} = H^a_{\ c}
\Gamma^{c}_{\ d} (H^{-1})^{d}_{\  b} + H^{a}_{\ c}
\de (H^{-1})^{c}_{\ b}, 
\end{equation}
which is equation (6) again. Thus, we have reduced the $G$-connection 
down to a $H$-connection (plus tensor parts). The matrix $r^{\alpha}_{\ b}$ 
is also called \textit{reducing matrix} \cite{6}. Note that you may 
see (8) as a gauge transformation (it is of the same form as (7), with 
$r^{\alpha}_{\ b}$ as a special element of $G$. Thus, in a sense, 
the reduction (8) is a gauge choice, but the gauge freedom is  not lost, 
the degrees of freedom have just been transferred to the reducing matrix. 
That is what we meant in the introduction by 
\textit{gauge parameterization}. 

Let us close this section by  demonstrating this formalism with an 
explicit example. For simplicity, we take the Poincar\'e group $ISO(3,1)$ 
as 
symmetry group $G$ which reduces to the Lorentz subgroup $H = SO(3,1)$.  
How can we find a suitable reducing matrix? Well, as we have pointed out, 
the reducing matrix is a special element of $G$ (i.e., a Poincar\'e 
transformation) that parameterizes the gauge degrees of freedom other than 
that of the Lorentz subgroup, i.e., the translations. Otherwise stated, 
the reducing matrix is a function of the coset parameters 
$\xi \in ISO(3,1)/SO(3,1)$. It is convenient to use a five dimensional 
matrix representation. We write the ten components of the Poincar\'e 
connection in the form 
\begin{equation}
\Gamma^{\bar A}_{\ \bar B} = 
\left( \begin{array}{ccc} \Gamma^{\bar a}_{\ \bar b} 
& \Gamma^{\bar a} \\ 0 & 0
\end{array} \right)
\end{equation} 
and a general Poincar\'e transformation as 
\begin{equation}
 P^{\bar A}_{\ \bar B} = \left( \begin{array}{cc} 
\Lambda^{\bar a}_{\ \bar b} &
 \xi^{\bar a} \\ 0 & 1 
\end{array} \right).
\end{equation} 
Clearly, the matrix that parameterizes the translations is given by 
\begin{equation}
r^{\bar A}_{\  B} = \left( \begin{array}{cc} 
\delta^{\bar a}_{  b} & \xi^{\bar a} 
\\ 0 & 1 
\end{array}\right).
\end{equation} 
We denote by $\bar A = (\bar a, \bar 5)$ 
the Poincar\'e group indices (instead of 
the greek ones, used above), in order to prevent confusion with the later 
on introduced $GL(R^4)$ indices, and by 
$A = ( a,  5)$ the reduced (Lorentz) 
indices. Straightforward application of (8) gives for the reduced 
connection 
\begin{equation}
\Gamma^A_{\ B} = \left( \begin{array}{cc} \Gamma^a_{\ b} & e^a \\ 0 & 0
\end{array} \right),
\end{equation}
where 
\begin{equation}
\Gamma^a_{\ b} = \delta^a_{\bar a}\  \delta^{\bar b}_b \Gamma^{\bar a}_{\ \bar
  b}\ \ \text{and} \ \ e^a = \delta^a_{\bar a}\ 
[ \Gamma^{\bar a} + \Gamma^{\bar a}_{\ \bar b}
\xi^b + \de \xi^{\bar a}]. 
\end{equation}
This is the well know expression for the Lorentz connection and the 
tetrad field (see \cite{17} or \cite{18}). 
It remains to show the transformation properties of the new 
fields and of the coset parameters $\xi^a$. This can be done using equation 
(3). 
In matrix notation, the short form  $\sigma' = \sigma(\xi')$ is to 
be interpreted as the fact that the transformed $r^{\bar A}_{\ B}$ (seen 
as element of the Poincar\'e group) is again a pure translation. 
Using the transformation law in the form (9), i.e., explicitly 
\begin{equation}
\tilde r^{\bar C}_{\  D} = \left( \begin{array}{cc} 
\Lambda^{\bar c}_{\ \bar a} &
a^{\bar c} \\ 0&1 \end{array} \right) \left( \begin{array}{cc} 
\delta^{\bar a}_{b} 
& \xi^{\bar a} \\ 0 &1 \end{array}\right) \left( \begin{array}{cc} 
(L^{-1})^b_{\ d}&0\\
0&1 \end{array}\right), 
\end{equation}
where the first matrix at the r.h.s. is a Poincar\'e transformation, 
and the third one a Lorentz transformation, a priori unrelated to each 
other, and then requiring that $\tilde r^{\bar C}_{\ D}$ is again a pure 
translation, i.e., of the form 
\begin{equation}
\tilde r^{\bar C}_{\ D} 
= \left( \begin{array}{cc} \delta^{\bar c}_{\ d}& \tilde \xi^{\bar c}\\
0&1 \end{array}\right),
\end{equation}
we find the condition $\Lambda^{\bar a}_{\ \bar b} 
= \delta^{\bar a}_a \delta^{b}_{\bar b} \ L^a_{\ b}$ (shortly $\Lambda=L$) 
and the transformation law 
\begin{equation}
\tilde \xi^a = \Lambda^a_{\ b} \xi^b + a^a.
\end{equation}
It is now easy to show that $\Gamma^a_{\ b}$ and $e^a$ in (15) transform  
indeed 
as Lorentz connection and Lorentz vector respectively. 

In our special example with $ISO(3,1)$ as symmetry group, there is 
actually nothing nonlinear in the realization. Indeed, the relation
$\Lambda = L$ allows us to  
 identify the barred indices with the unbarred ones in the 
four dimensional subspace, as we have done in (18).  
In more complex examples, like the 
de Sitter \cite{11} or the 
general linear \cite{19} group, the transformation law 
(3) leads to  a nonlinear dependence of the  residual Lorentz transformation 
on the Lorentz part of the transformation of the initial symmetry group. 
Also, 
the reduced connection will depend nonlinearly on the coset parameters. 
We will see those features in the next sections, in the context of 
spontaneously broken gauge theories.

\section{The de Sitter group}

In this section, we present the gravitational gauge theory of the 
de Sitter group, spontaneously broken down to the Lorentz group, 
as presented by Stelle and West \cite{11}. 
We do this not only because \cite{11} seems to be the only article where 
a (conventional) Higgs sector is actually 
explicitly written down, but also because 
the de Sitter case, although nontrivial, is nevertheless nearer 
to conventional broken gauge theories, like the electroweak model or 
the nonlinear sigma-model and therefore presents a good starting point 
for the consideration of other symmetry groups which may present some 
particularities. 

We start with a de Sitter (i.e., SO(4,1)) 
connection $\Gamma^A_{\ B}$ (antisymmetric 
when we raise one index with the de Sitter metric 
$\eta_{AB} = diag(1,-1,-1,-1,-1)$) and construct  the free Lagrangian 
density (or four form) $\mathcal L_0 $ depending,  ideally, on 
$\Gamma^A_{\ B}$ only. (We will later see that the free Lagrangian can 
also depend on the Higgs field, in contrast to conventional 
Yang-Mills theories.) Then, in order to break down the symmetry, we have 
to introduce a Higgs field $y^A$ carrying a de Sitter representation 
(in our case, $y^A$ is simply a de Sitter vector). Note that the $y^A$ are   
scalars under spacetime transformations. The Higgs potential $V(y^A)$ 
is chosen in a way that the groundstate $y^A(0)$ is invariant only under 
the Lorentz subgroup. 

The first step is to find a decent (de Sitter invariant) 
characterization of such a groundstate. 
This is not very hard in this case, we may take 
\begin{equation}
y^A(0)y_A(0) = -v^2,
\end{equation}
where $v^2$ is a constant. Let us suppose that we have a Higgs sector 
that leads to  (19) for the groundstate. (We will first consider the 
symmetry breaking and construct the Lagrangian afterwards.) The 
next step is to choose, from all the possible groundstates (19), 
the Lorentz invariant one 
\begin{equation}
y^A(0) = (0,0,0,0,v)
\end{equation}
and to expand $y^A$ in terms of new fields (which  vanish in the groundstate)
around $y^A(0)$. A convenient parameterization is given by 
\begin{equation}
y^A = \left( \begin{array}{c}  (\xi^a/\xi)(v+ \phi) \sin (\xi/v) \\
-(v+\phi) \cos (\xi/v) \end{array}\right), 
\end{equation}
where $\xi = \sqrt{- \eta_{ab}\xi^a\xi^b} = \sqrt{- \xi_a\xi^a}$.
Note that $y^Ay_A = - (v+ \phi)^2$, i.e., $\phi $ gives us a measure 
of how far we are from the groundstate. It is the residual 
Higgs field of the 
theory. (We call $y^A$ the (unbroken) Higgs field and $\phi$ the 
residual Higgs field, but occasionally refer to either of them as Higgs 
field.) Clearly, the remaining fields $\xi^a$ are pure gauge, i.e., 
they can be transformed away. Indeed, performing a de Sitter gauge
transformation with 
\begin{equation}\Lambda^A_{\ B} = 
\left( \begin{array}{cc} \delta^a_b +(\cos (\xi/v)\!+ 1)(\xi^a\xi_b/\xi^2) &
-  (\xi^a/\xi)\sin (\xi/v)  
\\&\\  -(\xi_b/\xi) \sin (\xi/v) &  -\cos (\xi/v)  
\end{array}\right) ,
\end{equation}
the Higgs field reduces to the form 
\begin{equation}
\tilde y^A = (0,0,0,0,v+\phi). 
\end{equation}
In order for our action to remain unaffected, we have to carry out the 
gauge transformation on all the fields, especially on the connection  
(we rescale the pseudo-translational part for dimensional reasons 
and take $e^a = v \Gamma^{a}_{\ 5}$, since we intend to interpret $e^a$ 
as tetrad field)
\begin{equation}
\Gamma^A_{\ B} = \left( \begin{array}{cc} \Gamma^a_{\ b} & \frac{1}{v}e^a \\ 
\frac{1}{v}e_b & 0
\end{array}\right), 
\end{equation}
which  under the transformation  
\begin{equation}
\tilde \Gamma^A_{\ B} 
= \Lambda^A_{\ C}\Gamma^C_{\ D}\Lambda^{\!\!-1D}_{\, \ \ B}
+ \Lambda^A_{\ C} \de \Lambda^{\!\!-1C}_{\, \ \ B}, 
\end{equation}
takes the form 
\begin{equation}
\tilde \Gamma^A_{\ B} = \left( \begin{array}{cc} \tilde \Gamma^a_{\ b} 
& \frac{1}{v}\tilde e^a \\ 
\frac{1}{v}\tilde e_b & 0 
\end{array} \right), 
\end{equation}
where 
\begin{eqnarray}
\tilde\Gamma^a_{\ b} &=& \Gamma^a_{\ b} + \frac{\cos (\xi/v) +1}{\xi^2}\  
(\Gamma^c_{\ b}\xi^a \xi_c + \Gamma^a_{\ c} \xi^c \xi_b)  
- \frac{\sin(\xi /v)}{v \xi}(\xi^a e_b - e^a \xi_b) \nonumber \\ &&
+ \frac{\cos(\xi/v) +1}{ \xi^2}
(\xi_b \de \xi^a  - \xi^a \de \xi_b), \\
\tilde e^a &=& - e^a \cos(\xi/v) 
+ v \Gamma^a_{\ c}\ \frac{\xi^c}{\xi} \sin(\xi/v) 
- \frac{1+\cos(\xi/v)}
{\xi^2}\  \xi^a e_c\xi^c 
 \nonumber \\ && + \frac{1+ \sin{\xi/v}}{\xi^3}\  \xi^c \de \xi_c\  
\xi^a + 
\frac{\sin(\xi/v)}{\xi} \de \xi^a. 
\end{eqnarray}
This result coincides with the one given in \cite{11},  after the 
replacement $\cos \rightarrow \cosh$ and $\sin \rightarrow \sinh$
and if one takes into account the difference in the sign convention. 
In \cite{11}, the signature is 
taken to be $\eta_{AB} = diag(-1,+1,+1,+1,-1)$, i.e., the symmetry group 
is $SO(3,2)$, the anti-de Sitter group, and the global sign convention is
opposite to ours. 

Note that this result was achieved without the use of the cumbersome 
exponential parameterization that is used  almost throughout the 
literature, including \cite{11}, and thereby, the painful manipulations   
associated with the use of the Campell-Hausdorff formula could be 
avoided. 

If we compare with the last section, we recognize in (22) the reducing 
matrix, that is used to produce the nonlinear Lorentz connection (27) 
and the tensor (or rather vector) part (28). We also see that the coset 
parameters
$\xi^a$ appear now as the Nambu-Goldstone bosons of the theory. 
They are removed (gauged away) from the Higgs field and absorbed by 
the connection. 

The remaining step is the construction of the Lagrangian. The appropriate Higgs
sector has been constructed by Stelle and West. Introduce the following 
tensor:
\begin{equation}
g_{ik} = \frac{v^2}{y^Cy_C}[\frac{1}{y^D y_D}(y^A \De_i y_A)
(y^B \De_k y_B)-\De_i y^A \De_k y_A], 
\end{equation}
 where  $\De_i y^A = y^A_{,i} + \Gamma^A_{\ Bi}y^B$. Clearly, 
$g_{ik}$ is  de Sitter gauge invariant. In the gauge (23), it takes the simple 
form 
\begin{equation}
g_{ik} = \tilde e^a_i \tilde e^b_k \eta_{ab}. 
\end{equation}
We can now construct the following Higgs Lagrangian density: 
\begin{equation}
\mathcal L_{higgs} = \sqrt{-g}[- \frac{1}{2y^Cy_C}(y^B \De_iy_B)(y^C \De_k
y_C)g^{ik}-V(y)], 
\end{equation}
with the Higgs potential 
\begin{equation}
V  =  \frac{m}{2} y^Ay_A + \frac{\lambda}{4} (y^Ay_A)^2, 
\end{equation}
with $m$ and $\lambda$ positive constants. Obviously, the groundstate is
characterized by the conditions 
\begin{equation} 
y_Ay^A = - \frac{m}{\lambda} = - v^2, 
\end{equation} 
and 
\begin{equation}
y^A\De_i y_A = 0. 
\end{equation}
Note that this last condition somewhat differs from the usual case. One would
have expected a kinetic term of the  
form $(D_i y^A)^2$ and a groundstate 
condition $D_i y^A = 0$. This however would destroy the possibility of 
interpreting  
$\tilde e^a_m$ as tetrad field (and  $g_{ik}$ as metric), because then, in 
the groundstate the tetrad would vanish and certainly not be invertible. 

In the gauge (23), the Higgs Lagrangian reduces to 
\begin{equation}
\mathcal L_{higgs} =  \sqrt{-g}[ \frac{1}{2}\phi_{,i}\phi^{,i}+ \frac{m}{2}
(v+\phi)^2 - \frac{\lambda}{4}(v + \phi)^4], 
 \end{equation}
which is of the conventional form. 

It remains to complete the theory with the Lagrangian for the gravitational
field itself. In conventional theories, the Higgs field does not appear in 
the Yang-Mills sector. In gravity, this is not always possible, as we will
see, because it is rather difficult to construct a Lagrangian four form using 
exclusively the gauge potentials. However, in the de Sitter case, Stelle and 
West 
have presented such a Lagrangian. It is of the form 
\begin{equation}
\mathcal L_0\!\! =\!\!
\sqrt{\!-(F^{AB}\!\wedge F^{CD} \epsilon_{ABCDE})(F^{FG}\!\wedge F^{HI} 
\epsilon_{FGHIJ})\eta^{EI}},
\end{equation}
where $F^{AB}$ is the curvature tensor of the de Sitter group which has the
following components
\begin{equation}
F^{ab} = R^{ab} + \frac{1}{v^2}(e^a \wedge e^b),\ \ F^{a5} = \frac{1}{v}(\de
e^a + \Gamma^a_{\ b} \wedge e^b). 
\end{equation}
Going over to the  gauge (23) (which means simply replacing $\Gamma, e$ by
$\tilde \Gamma, \tilde e$), 
where $\tilde \Gamma^a_{\ b}$ and $\tilde e^a$ are now the Lorentz 
connection and the tetrad, we recognize (dropping the $\ \tilde{}\  $ for
convenience) 
in $R^{ab}$ the Lorentz curvature and 
in $F^{a5}$ the torsion 2-form $T^a$. 

There is one case where we can simplify (36). If the torsion vanishes, we
simply get,  omitting in the second line the Gauss-Bonnet divergence term  
\begin{eqnarray}
\mathcal L_0 &= & -F^{ab}\wedge F^{cd} \epsilon_{abcd} \nonumber \\ 
&=& -2 R^{ab} \wedge e^c \wedge e^d \epsilon_{abcd} - e^a \wedge e^b \wedge e^c
\wedge e^d \epsilon_{abcd} \nonumber \\
&=&  e\  \frac{8}{v^2}(R - \frac{3}{v^2}),  
\end{eqnarray}
which is general relativity (in the first order formalism) 
with a cosmological constant. Unless some matter field is coupled to
the Higgs field $\phi$ and to the Lorentz connection (as is the case with
fermion fields), the theory remains in its groundstate $\phi = 0$ 
and $T^a = 0$. 
The general case with non-vanishing torsion is difficult to handle with the 
non polynomial Lagrangian. 

An alternative Lagrangian has also been proposed by Stelle and West, namely 
\begin{equation}
\mathcal L_0 =   F^{AB}\wedge F^{CD} \epsilon_{ABCDE}\ y^E/v. 
\end{equation}
As before, using (37), you will find (in the gauge (23)), a curvature scalar
term, 
a Gauss-Bonnet term and a cosmological constant term, each of them appearing,
however,  
with a factor $(v+ \phi)$. Rescaling the scalar field by $1/v$ to make it 
dimensionless, 
and shifting it by one (to get a groundstate $\phi = 1$ instead of $\phi =
0$), the theory reveals itself to be 
very similar to Brans-Dicke theory (in a Riemann-Cartan 
framework). Note however that we cannot omit the Gauss-Bonnet term anymore. 
This term, of second order in the curvature, will eventually lead to  
dynamical 
torsion fields (in addition to those induced by the scalar field, 
see \cite{11}) in the presence
of spinning matter fields.

\section{The Poincare group}

The main difference between the de Sitter and the Poincar\'e group is the fact 
that the latter is not semi-simple. As a result, there is no Cartan metric 
available. This makes it rather difficult to construct invariants. 

Let us begin by the introduction of a Higgs field $y^a$, carrying a 
(vector) representation of the Poincar\'e group, i.e., transforming 
as 
\begin{equation}
y^a \rightarrow \Lambda^a_{\ b} y^b + a^a
\end{equation} 
under a Poincar\'e transformation $(\Lambda^a_{\ b}, a^a)$. 
For convenience, we 
do not distinguish notationally between a Lorentz and a Poincar\'e index (as
we did in section 2). 

The next step is to find a groundstate $y^a(0)$ that is Lorentz invariant, 
but breaks
the Poincar\'e invariance. For the groundstate to be Lorentz invariant,
it can only be characterized by  a Lorentz invariant relation 
$\eta_{ab}\ y^a(0) y^b(0) = v^2$. Clearly, the only case where we have a 
Lorentz invariant solution to this is for $v^2 = 0$. It is the state 
\begin{equation}
y^a(0) = 0. 
\end{equation}
This is indeed Lorentz invariant and breaks the Poincar\'e
invariance. 

To hold track with the de Sitter case, equation (41) corresponds to 
equation (20). We have
now to parameterize a general state in terms of the groundstate, and write 
\begin{equation}
 y^a = \xi^a, 
\end{equation}
which is the counterpart of equation (21). This looks quite trivial, but it is
the most general parameterization there is. Not surprisingly, it is pure gauge
(as is any Poincar\'e vector). Otherwise stated, $y^a$ is expressed entirely 
in terms of the Nambu-Goldstone bosons. There is no residual Higgs field in 
the theory. 
The transformation corresponding to the
pseudo-translation (22) is now a pure translation (it is the $r^A_{\ \bar B}$
of section 2). Transforming the Higgs field and the connection, we find 
\begin{equation}
\tilde y^a = 0, \ \ \tilde \Gamma^a_{\ b} = \Gamma^a_{\ b},\  \ \tilde
\Gamma^a  \equiv e^a = \Gamma^a + \de \xi^a + \Gamma^a_{\ b} \xi^b.
\end{equation}
This is to be compared with (27) and (28). 

We go on constructing the Higgs Lagrangian. What kind of a Higgs 
potential allows for a groundstate $y^a(0) = 0$? Well, the most 
simple solution is certainly $V(y) = 0$. This is not only short and 
elegant, but seems to be also the unique possible choice. As mentioned 
above, in Poincar\'e gauge theory, there are rather few objects to 
construct invariants. The Minkowski metic $\eta_{ab}$, although not 
related to a Cartan metric, nevertheless constitutes a natural ingredient of 
the theory, since it appears explicitly in the structure constants of the 
group algebra. It is therefore not unnatural to simply declare it as
\textit{invariant} under Poincar\'e transformations. You can also see it 
simply as a constant matrix. Similarly, you may see $\epsilon_{abcd}$ as 
Poincar\'e invariant. 

With these objects, however, and the Poincar\'e vector 
$y^a$, the only scalar density you can construct is given by 
\begin{equation}
\sqrt{-g}\ V(y) =  \sqrt{-g}\ \lambda, 
\end{equation}
with a constant $\lambda$ and with 
\begin{equation}
g_{ik} = (\De^{PG}_i y^a )(\De^{PG}_k y^b )\eta_{ab}, 
\end{equation}
where $\De_i^{PG} $ denotes  
the Poincar\'e covariant derivative, $\De^{PG}_i y^a = 
y^a_{,i} + \Gamma^a_{\ bi}y^b + \Gamma^a_i$. Note that this is not really 
a covariant derivative in the sense that it transforms  in the same 
way as $y^a$. It has however the nice property that its vanishing in one 
gauge assures the vanishing in any gauge. In other words, it is a Lorentz 
vector. The metric $g_{ik}$ is of course gauge invariant. 

In (44),  the potential $V$ actually consists  
of an (almost) trivial constant $\lambda$. The only reason for which 
it is not trivial 
 are the fields that appear inside the metric $g_{ik}$ in the factor
 $\sqrt{-g}$.. 
But in view of this, 
and considering (45), we should rather consider (44) as part of the 
kinetic term of the Higgs Lagrangian.  We therefore 
conclude that  $V(y) = 0$. 

As to the kinetic part, we already found one candidate with (44), namely 
$\sqrt{-g} \lambda$. Other ideas, like 

$\epsilon_{abcd}
\De^{PG} y^a \wedge \De^{PG} y^b \wedge \De^{PG} y^c \wedge \De^{PG} y^d $ 
or 
$\sqrt{-g} g^{ik} \De^{PG}_i y^a \De_k^{PG} y_a$ 

are only different forms 
or the same expression. No other expression containing 
only first derivatives of $y^a$ can be 
formed with the ingredients we have at our disposal. Indeed, the next
most simple candidate is the Riemannian curvature scalar for the 
metric $g_{ik}$. This contains higher derivatives of $y^a$ (in the metric 
derivatives) and moreover, the Riemannian curvature can also be formed 
as a part of $\mathcal L_0$, as we know from traditional 
Poincar\'e gauge theory. It is then written as the sum of the teleparallel 
Lagrangian and the (Lorentz) curvature scalar.  
 
The total, most general Higgs Lagrangian therefore has the form 
\begin{equation}
\mathcal L_{higgs} = \sqrt{-g}\ \frac{\lambda}{4} 
 g^{ik} \De_i^{PG}y^a \De_k^{PG} y_a 
= \sqrt{-g}\ \lambda. 
\end{equation}
 In its first form, this looks like a conventional kinetic term for 
the Higgs field $y^a$, but one should not forget that during the 
variation, the metric has to be taken into account too. In the gauge 
(43), we simply get (omitting the $\ \tilde{}\ $) 
\begin{equation}
\mathcal L_{higgs} = e\ \lambda, 
\end{equation}
where $e$ denotes the determinant of $e^a_m$. Summarizing, the Higgs potential 
is zero and the kinetic Higgs sector corresponds to a cosmological constant. 
This explains why apparently, in Poincar\'e gauge theory, nobody included 
a Higgs sector. It was already there, but we  did not recognize 
it as such. 

It remains to complete the theory by constructing the gravitational sector. 
The only covariant quantity we can construct out of the Poincar\'e 
connection (we use again a five dimensional representation in the 
form (11)) is the Yang-Mills tensor 
\begin{equation}
F^A_{\ B} = \left( \begin{array}{cc} R^a_{\ b} & \tau^a \\0&0 \end{array}
\right)
\end{equation}
where $R^a_{\ b}$ is the Lorentz curvature and $\tau^a = \de \Gamma^a
+ \Gamma^a_{\ b}\wedge \Gamma^b$. You can also check that the
$\epsilon_{ABCDE}$ is again a tensor density under (12). With these 
objects alone, it is quite difficult, although not impossible, to construct 
Lagrangians. To be complete, we 
give an example:
\begin{equation}
\mathcal L_0 = \sqrt{(F^A_{\ E}\wedge F^B_{\ F})(F^C_{\ G}\wedge F^D_{\ H})
\epsilon_{ABCDI} \epsilon^{EFGHI}}. 
\end{equation}
We do not know if it is possible to reduce this expression to 
something more easy to handle, or even to something useful, but it 
shows that there are, at least in principle, actions that are invariant 
under the Poincar\'e group, without the help of additional structures. 
 
You may also look for actions formed out of the four dimensional parts 
$R^a_{\ b}$ and $\tau^a$. This has the advantage that, in addition to 
 $\epsilon_{abcd}$, you also have $\eta_{ab}$ as invariant. However, 
no obvious scalar can be constructed with $\tau^a $, due to 
its non homogeneous 
behavior under translations, and quantities constructed with the curvature, 
like $R^{ab} \wedge R_{ab}$ or $R^{ab}\wedge R^{cd} \epsilon_{abcd}$ turn 
out to be total derivatives. (Of course, there is also a solution to that: 
Take the square of the first, add the square of the second and take the 
square root. The result is certainly not a total derivative.) 

In order to construct more conventional, polynomial, Lagrangians, we see that 
the only way is to use the Higgs field also in the gravitational part of 
the Lagrangian. The Einstein-Cartan Lagrangian, for instance, is written as 
\begin{equation}
\mathcal L_0 = \De^{PG} y^a  \wedge \De^{PG} y^b \wedge R^{cd} \epsilon_{abcd}
\end{equation}
which is gauge invariant and reduces in the gauge (43) to 
\begin{equation}
\mathcal L_0 = e^a  \wedge e^b \wedge R^{cd} \epsilon_{abcd}. 
\end{equation}
More generally, if we  introduce the fields 
\begin{equation}
E^a = \De^{PG} y^a\ \  \text{and}\ \   T^a = \de E^a + \Gamma^a_{\ b}\wedge 
E^b 
\end{equation}
then any Lagrangian 
constructed out of  $E^a, R^{ab}$ and $T^a$
will be Poincar\'e invariant. These quantities are  Lorentz tensors (i.e., 
invariant under translations). $E^a$
plays the role of the tetrad field, 
as is clear in  the gauge (43), where it 
reduces to $E^a = e^a$. In the same gauge, $T^a$ reduces to the 
torsion tensor.
In this way, you get 
the whole family of conventional Poincar\'e gauge theory Lagrangians.  

More generally, one can argue that no Lagrangian can be constructed 
without the help of $y^a$ that contains the translational 
gauge field $\Gamma^a$. You see this as follows: If the field $y^a$ is 
contained in $\mathcal L_0$, then it appears necessarily only in the 
combination $\De^{PG} y^a$. This however, in the specific 
gauge  $y^a = 0$, reduces to $\Gamma^a$. The other way around, if 
the Lagrangian depends only on $\Gamma^a_{\ b}$ and $\Gamma^a$, you can 
interpret $\Gamma^a$  as $\De^{PG} y^a$ in the gauge $y^a = 0$, since the 
Lagrangian is gauge invariant. This would mean that, in addition to 
its Poincar\'e gauge invariance, the Lagrangian would have to be 
invariant under the \textit{replacement}  $\Gamma^a \rightarrow \De^{PG} y^a$, 
since else, it  can not be gauge invariant with and without the $y^a$ terms. 
Clearly, this is only possible when $\Gamma^a$ is not contained at 
all in $\mathcal L_0$. Note that this is also the case with (49), which
reduces immediately to its four dimensional parts,  since $F^A_{\ 5} = 0$. 

However, if $\mathcal L_0$ does not depend on $\Gamma^a$, i.e., on $e^a$ 
in the final gauge, then we will not have an Einstein equation,  which 
means that the energy-stress tensor of the matter fields will have to 
vanish. Such Lagrangians can therfore not be considered as physically 
acceptable. As a result, the Higgs field $y^a$ is a necessary ingredient 
of Poincar\'e gauge theory, apart from its role in the  symmetry breaking 
process. 

Before we close this section, let us make some remarks about the groundstate
of the theory. We have concluded earlier that the groundstate has to be 
described by $y^a = 0$, because it is the only choice that is Lorentz 
invariant. However, the groundstate can not be guessed simply from the 
form of the potential or from other arguments, it has to be the result of 
the field equations for the Higgs field. If you look at (46), you might 
get the impression that the groundstate is characterized 
by $\De^{PG}_i y^a = 0$. 
This would be rather unfortunate, since this is just the quantity that 
is to be used as tetrad field. We should therefore check the situation 
more carefully. 

Let us first consider the (conventional, but in gravity unlikely) 
case where the Lagrangian $\mathcal L_0$ does not depend  on the Higgs field.  
Then, in the absence of matter field that couple to $y^a$, the field 
equation for $y^a$ is easily derived by varying (46) 
\begin{equation}
T^b_{\ ik} E^i_a E^k_b = 0, 
\end{equation}
or simply, transforming tangent space indices into spacetime indices 
using the tetrad, $T^l_{\ ml} = 0$. The groundstate is thus characterized 
by equation (53), which is the analogue of equation (34) in the de Sitter 
case. This is clearly a Poincar\'e covariant relation (meaning that, if 
it holds in one gauge, then in any; not meaning that the left hand 
side is a Poincar\'e vector) and you may now go 
on and choose the state $y^a(0)$ in Lorentz invariant way. 

In the usual case, the Lagrangian $\mathcal L_0$, and also 
any matter Lagrangian $\mathcal L_m$ will depend on $y^a$ through 
the tetrad field $E^a = \De^{PG} y^a$. In this case, it is easy to see 
that the equation arising from variation with respect to  $y^a$ is 
identically satisfied if the Einstein equation (i.e., the equation for $E^a$, 
or, equivalently, for $\Gamma^a$) 
is satisfied. Indeed, the $y^a$ equation is just the covariant derivative of 
the $E^a$ equation, see \cite{17}. Therefore, no additional condition 
arises on the tetrad field that would exclude its invertibility. 

In practice, the usual procedure is to first gauge the Nambu-Goldstone 
bosons away and 
to vary the action afterwards. Then, since $y^a$ was pure gauge, the 
Higgs field does not appear anymore, and the only thing that reminds us 
of it is the cosmological constant (i.e., a mass term for the field $e^a$, 
the field corresponding to the broken gauge symmetry). We see that the 
fact that $y^a$ is pure gauge is directly related to the fact that 
its field equation is identically satisfied and that 
after the symmetry breaking, no Higgs field remains in the theory. 

There is only one question left: Can we choose $\lambda = 0$? Well, 
from a practical point of view, there is no problem with this. Theories 
without cosmological constant are quite commonly accepted. Choosing $\lambda 
= 0$ erases the last trace of the Higgs field and leaves $y^a$ without 
kinetic term. In the spirit of conventional theories with dynamical 
symmetry breaking, even though $y^a$ is 
of pure gauge nature, and we therefore can live without this term, 
one should not omit it. As kinetic term for the field $y^a$ (look at 
it in the first form of equation (46)), one might even argue that 
the \textit{natural} choice would correspond to $\lambda > 0$, 
i.e., to a positive cosmological constant.

\section{The affine group}

Considering the affine group as symmetry group of the gravitational 
interaction leads to the so called metric affine theory. 
We refer to \cite{20} for a detailed presentation of its features. 
The nonlinear realization with the Lorentz group as stability subgroup, 
as well as references to earlier attempts can be found in \cite{19}.  

To the affine group correspond  20 gauge fields $\Gamma^{\alpha}_{\ \beta}$ 
and $\Gamma^{\alpha}$, $\alpha, \beta = 0,1,2,3$. With the translational 
part $\Gamma^{\alpha}$, one deals exactly as in the Poincar\'e case, i.e., 
introduces a Higgs 
field $y^{\alpha}$, transforming as affine vector, 
then defines $E^{\alpha} = \Gamma^{\alpha} 
+ \de y^{\alpha} + \Gamma^{\alpha}_{\ \beta}y^{\beta}$ and so on. After 
gauging away the Nambu-Goldstone bosons, we have $E^{\alpha} = \Gamma^{\alpha} 
\equiv e^{\alpha}$. We are then left with the 20 fields $\Gamma^{\alpha}_{\
  \beta}$ and $e^{\alpha}$,  a $GL(R^4)$-connection and a
$GL(R^4)$-vector respectively. 
  
In this section, we concentrate on the reduction of  $GL(R^4)$ 
(shortly $GL$) to $O(3,1)$. Let us briefly review the nonlinear 
realization process. As in section 2, we introduce a reducing 
matrix $r^{\alpha}_{\ b}$ with a mixed transformation behavior given 
by equation (9), where $H^a_{\ b}$ is now a Lorentz  
and $G^{\alpha}_{\ \beta}$ a $GL$ transformation. The Lorentz connection 
is then given by (8). 

What is a suitable reducing matrix? Recall that 
$r^{\alpha}_{\ a}$ is itself an element of $GL$. In the 
 Poincar\'e case, the reducing matrix was a pure translation. Similarly, in 
this case, it will be a general linear transformation that does 
\textit{not contain} 
a Lorentz rotation. We know that infinitesimal Lorentz transformations are 
antisymmetric. Antisymmetric with respect to the Lorentz metric, more 
 precisely. Because, if we take any (infinitesimal) 
general linear transformation $\epsilon^{\alpha}_{\ \beta}$, and look 
at a general metric tensor $g^{\alpha \beta}$, then this metric 
will be invariant if we have $\epsilon^{\alpha}_{\ \gamma} g^{\gamma \beta}
=- \epsilon^{\beta}_{\ \gamma}g^{\alpha \gamma}$, 
i.e., if $\epsilon^{\alpha \beta}$ is antisymmetric where the second index has 
been raised with $g^{\alpha \beta}$. It is a Lorentz transformation, by 
definition, only in the case where 
 the invariant metric is the Minkowski metric. In that case, we say that  
$\epsilon^{\alpha}_{\ \beta}$ is antisymmetric with respect to the 
Minkowski (or Lorentz) metric. In \cite{7}, the expressions pseudo-symmetric 
and pseudo-antisymmetric are used in this sense.

The infinitesimal Lorentz transformations being antisymmetric in the 
above sense, the infinitesimal form of our reducing matrices have to 
be symmetric. Exponentiating a symmetric matrix leads to a symmetric matrix. 
Therefore, we finally have the following characterization: 
\begin{equation}
r^{\alpha}_{\ a}\eta^{a\beta} = r^{\beta}_{\ a}\eta^{a\alpha}.
\end{equation}
This reduces the independent degrees of freedom contained in 
$r^{\alpha}_{\ a}$ 
to 10, as many as the parameterized gauge degrees of freedom. 

The appearance of the Minkowski metric is unavoidable at this stage, 
it is the result of having chosen  a specific stability subgroup.
If we take $diag(1,1,1,1)$ instead of $\eta_{ab}$, we will end up 
with an $O(4)$ symmetry. Unfortunately, with the exception of \cite{7}, 
little attention has been paid to this crucial point. 
One usually splits the $GL$ generators $G^{\alpha}_{\ \beta}$ into  
a \textit{Lorentz} and a \textit{symmetric} part 
$L^{\alpha}_{\ \beta} + S^{\alpha}_{\ \beta}$, without mentioning anything 
about a metric. The truth is, however, 
that you cannot split in this way the generators \textit{before} 
the introduction of the Minkowski metric. This can  already be seen from 
the concrete representation with the $GL$-generators taken in the form  
$x^k\partial_i$. You cannot extract the angular momentum generators 
from this, without the use of a metric. The geometric 
reason for this is the fact that in dimensions higher than three, 
rotations take place not \textit{around an axis}, but \textit{in a surface}, 
and therefore you have to know the geometry of that surface.  

The requirement, that after a transformation (9), the reducing matrix 
remains symmetric in the sense of (54) leads to a nonlinear relation 
between the Lorentz part of the $GL$ transformation $G^{\alpha}_{\ \beta}$
and the Lorentz transformation $H^a_{\ b}$ in (9) (see \cite{19}). 

You can also define a metric through $g_{\alpha \beta} = r^a_{\ \alpha} 
r^b_{\ \beta} \eta_{ab}$, transforming as tensor under the $GL$ group, 
but this seems rather unnatural at this point since our goal is just 
the contrary, i.e., to reduce $GL$ quantities to Lorentz quantities. 

We move  on to the construction of a theory with dynamical breakdown 
of the symmetry group $GL$ to the Lorentz stability subgroup $O(3,1)$.    
   
Together with the fields $e^{\alpha}$ and $\Gamma^{\alpha}_{\ \beta}$ 
(recall that we consider the translational gauge symmetry to be  
already broken), 
we introduce 10 Higgs fields (as many as the gauge degrees of freedom we want
to break) carrying a $GL$ representation, in the form 
of a symmetric tensor $g_{\alpha\beta}$. We can, at this stage, also 
introduce a spacetime metric through 
\begin{equation}
g_{ik} = e^{\alpha}_i e^{\beta}_k g_{\alpha\beta}. 
\end{equation}
We now look for a Higgs potential that is supposed to lead to a  
 Lorentz invariant groundstate, but breaks the  $GL$ symmetry. 
The following considerations will help us to construct  it. 
Recall the fact that $GL$, as opposed to 
the Lorentz group, is a very strong group, in the sense that we can 
diagonalize any symmetric matrix. More precisely, we can bring 
$g_{\alpha \beta}$ into one of the four forms 
$diag(\pm 1, \pm 1, -1,-1)$, depending on the signature of $g_{\alpha\beta}$
 (see \cite{20}). 
Note that we are not interested in the global sign of the 
metric, which is physically not relevant (different conventions are 
used in the literature with equivalent results).
 Therefore, if the groundstate turns out to have a specific 
signature, then the stability subgroup is fixed to the corresponding 
rotation group $O(3,1),\  O(2,2)$ or $O(4)$. It seems quite 
impossible to write down a Higgs potential depending on the  signature 
(expressed in terms of the $g_{\alpha \beta}$ components), but fortunately, 
there is one case where it is enough to know the sign of the determinant
in order to conclude for the signature. This is the case where 
 the first two entries in $diag(\pm 1, \pm 1,-1,-1)$ are of opposite 
sign. Only then will the determinant be negative. The determinant is 
not an invariant, but it does not change sign under a $GL$ transformation. 
This is the clue to the construction of the Higgs potential. All we 
have to do is to construct a Lagrangian that leads to a groundstate 
characterization  $\det g_{\alpha\beta} < 0$. 

This, however, turns out to be more difficult than it seems. 
The most simple scalar density that is invariant under $GL$ 
and could eventually serve our purpose is the following 
\[
\sqrt{|g_{ik}|}\ V(g_{\alpha\beta}) 
= |e|\ \sqrt{|\det g_{\alpha\beta}| + \det g_{\alpha\beta}}. 
\]
Indeed, variation with respect to $g_{\alpha\beta}$ leads to 
\begin{equation}
\delta(\sqrt{|g_{ik}|}\ V(g_{\alpha\beta})) 
= - |e|\sqrt{|\det g_{\alpha\beta}|+ \det g_{\alpha\beta}}\ \ g_{\alpha\beta}
\ \delta g^{\alpha\beta}, 
\end{equation}
and consequently, the groundstate is seemingly characterized by the condition
\begin{displaymath}
\det g_{\alpha \beta} < 0. 
\end{displaymath}
One could complete the Higgs sector in the following way 
\begin{displaymath}
\mathcal L_{higgs} = \sqrt{|g_{ik}|}\ \left[ g^{ik} \De_i g_{\alpha\beta}
\De_k g^{\alpha \beta}   - V(g_{\alpha\beta})        \right]  
\end{displaymath} 
It seems as if, as long as the non-metricity $\De g_{\alpha\beta}$ is 
zero, the theory would remain in its groundstate, i.e., the metric 
signature would be Lorentzian. This, however, is an illusion. 

The problem with (56) is, that we will never get this equation 
in an isolated form, even if $\De g_{\alpha\beta} = 0$. You can 
directly check that an equivalent equation will be obtained 
from the variation with respect to $e^{\alpha}$. Moreover, it is 
a general result  \cite{20}  
that in metric affine theory, one of the 
two equations arising from the variation with respect to the tetrad 
and to the metric respectively, is redundant (under the assumption 
that the matter equations and the $\Gamma^{\alpha}_{\ \beta}$ 
equations hold). 
The reason for this is that you can always gauge the metric into 
a non-dynamical, constant matrix 
(not necessarily of Minkowskian signature), or 
alternatively, you can gauge $e^{\alpha}_i$ into $\delta^{\alpha}_i$. 
Thus, one of both fields can be seen as pure gauge. 

After these considerations, it is clear that the expression (56) will 
arise in both the $e^{\alpha}$ equation (i.e., ultimately, in the 
Einstein equation) and in the
$g_{\alpha\beta}$ equation. There will be no independent equation for 
$g_{\alpha\beta}$. The same will be true for whatever Higgs Lagrangian 
$\mathcal L_{higgs} (g)$ one might come up with.  

Before we continue, let us note that the above argument also leads to 
the conclusion, that no Lagrangian can be constructed for the free 
gravitational fields without the help of 
$g_{\alpha\beta}$, apart from those that do not contain  $e^{\alpha}$, 
i.e., those who require a vanishing of the stress-energy tensor of the 
matter fields. Because, if $\mathcal L_0$ does not depend on the metric, 
its $e^{\alpha}$ equation too  has to be identically satisfied (with 
the help of the other equations). 

On the other hand, if we start with the symmetry group $SL^{\pm}$ 
(the subgroup of $GL$ with $|\det G|= 1$ for $G \in GL$), this 
problem should not occur, since we cannot, in general, gauge $e^{\alpha}_i$ 
into $\delta^{\alpha}_i$. Indeed, under $SL^{\pm}$, we have an additional 
invariant given by $\det g_{\alpha\beta}$, and we can construct 
the following Higgs potential 
\begin{equation}
\sqrt{|\det g_{ik}|}\ V = \sqrt{|g_{ik}|}\ 
[(\det g_{\alpha\beta}) + \frac{1}{2} (\det g_{\alpha\beta})^2]. 
\end{equation}
The only part of the equation  resulting  from the variation 
with respect to $g_{\alpha\beta}$ that will be  independent from 
the other equations (especially from the $e^{\alpha}$ equation), is 
the part coming from the $\det g_{\alpha\beta}$ variation inside the 
brackets. Therefore, we are let to 
\begin{equation}
\det g_{\alpha\beta} = -1. 
\end{equation}  
You can now add a dynamical sector for the Higgs field $g_{\alpha\beta}$. 
The groundstate condition (58) clearly breaks the symmetry 
from $SL^{\pm}$ down to $O(3,1)$. As long as the system 
remains in its groundstate, the metric will be related by an $SL^{\pm}$ 
gauge transformation to the Minkowski metric. We see that the 
introduction of additional (mani)fields is really not needed at this 
point. As opposed to the presentation of Hehl et al. 
in \cite{20}, in our approach, the nontrivial step is not the 
reduction from $SL^{\pm}$ to $O(3,1)$, but rather the reduction  
$GL \rightarrow SL^{\pm}$. 

How can we generalize (57) to the case of the $GL$ group? Well, we 
have already argued that we will never get an equation for $g_{\alpha\beta}$ 
that is independent of the rest of the equations, including the matter 
field equations. Therefore, the necessary step is to include a new 
matter field that serves the purpose to break the equivalence 
between the other two equations. In other words, if we introduce 
an additional Higgs field $\phi$, then we have the choice to 
consider either the set $(\Gamma^{\alpha}_{\ \beta}, e^{\alpha},\phi, \psi)$, 
where $\psi$ summarizes the other matter fields eventually present, 
or alternatively the set $(\Gamma^{\alpha}_{\ \beta}, e^{\alpha}, 
g_{\alpha\beta}, \psi)$ as independent variables. For this to work, 
it is clear that $\phi$ should be of pure gauge nature, that is, it should 
be possible to gauge it into something non-dynamical. After doing this, 
the gauge is 
fixed and we cannot trivialize anymore the tetrad or the metric, which 
will both take an independent status. Or the other way around, fixing 
the gauge by choosing the tetrad (or the metric) to be non-dynamical, 
the field $\phi$ becomes an independent dynamical quantity. 

Clearly, $\phi$ cannot be a $GL$ invariant. The next most simple candidate 
is a scalar density. This is also suggested by equation (57). It 
can easily be generalized if we choose $\phi$ to transform in a way that 
 $\phi^2 \det g_{\alpha\beta}$ is an $GL$ invariant. In other words, 
$\phi$ has to be a scalar density of weight $+1$ under $GL$ (and an ordinary  
scalar under spacetime transformations). Such a \textit{dilaton} field 
has also been used in \cite{20} in a similar approach. Note, however, 
that in contrast to \cite{20}, where 
the dilaton transforms as density under 
\textit{conformal} transformations, whereas it is a scalar under 
$GL$ transformations, we do not consider conformal transformations 
here. 

Finally, we are ready to propose the following Higgs Lagrangian: 
\begin{eqnarray}
\mathcal L_{higgs}  &=&  \sqrt{|g_{ik}|}\ 
\left[
\phi^2 (\det g_{\alpha\beta}) + \frac{1}{2} \phi^4 (\det g_{\alpha\beta})^2
+  \frac{1}{2} 
(\det g_{\alpha\beta})\ g^{ik} \De_i \phi \De_k \phi \right. \nonumber \\
&& \ \ \ \ \ \ \ \ \ \ \left.
+ g^{ik} \De_i g_{\alpha\beta}\De_k g^{\alpha\beta}\right]. 
\end{eqnarray}
The covariant derivative of $\phi$ is defined by 
\begin{equation} 
\De \phi = \de \phi + \Gamma^{\alpha}_{\ \alpha} \phi. 
\end{equation}
You can check that $\De \phi$ is again a scalar density of weight one. 
The metric determinant plays the role of the metric in the space of 
scalar densities, i.e., from two scalar densities $\phi, \psi$, you 
form the scalar $\phi \cdot \psi = (\det g_{\alpha\beta}) \phi \psi$. 
In this sense, the first two terms in (59) are of the usual form 
of a Higgs potential $\phi^2 + \frac{1}{2} \phi^4$ for the field $\phi$. 
(We have not included possible coupling constants in (59), since 
we are only interested in the main features of the symmetry breaking 
and not in specific models.)

The field equation for $\phi$ leads to the groundstate condition 
\begin{equation}
(g + g^2 \phi^2 ) \phi = 0, 
\end{equation}
where $g=\det g_{\alpha\beta} $, which, under the assumption $\phi, g\neq 0$, 
finally leads to 
\begin{equation}
\det g_{\alpha\beta} < 0. 
\end{equation}
This, as we have argued, leads uniquely to the Lorentz signature of the 
metric. The same signature, through (55), is passed on to the 
physical spacetime metric.

Before we comment on the above approach, let us complete our procedure 
in a way similar to the previous sections. The gravitational Lagrangian 
may be taken as  
\begin{equation}
\mathcal L_0 = \sqrt{|\det g_{\alpha\beta}|}\ 
\frac{e}{|e|}\ 
e^{\alpha}\wedge  e^{\beta} \wedge R^{\gamma}_{\ \kappa}\ \epsilon_{\alpha
  \beta \gamma \delta}\ g^{\delta \kappa}.    
\end{equation}
This is the direct generalization of the Einstein-Hilbert action to the 
metric affine framework. More general candidates can be found in \cite{20}. 
The factor $e/|e|$ is necessary in order to insure invariance 
under a $GL$ transformation with negative determinant, although it 
does not really change the field equations. From the groundstates 
characterized by (61), we choose 
\begin{equation}
|\phi(0)| = 1,\ \ g_{\alpha \beta}(0) = \eta_{\alpha\beta}.
\end{equation}
Any other choice, with the exception of $\phi = 0$ or $g= 0$ is related 
to this one by a $GL$ transformation. The first equation in (64) reduces 
the symmetry from $GL$ to $SL^{\pm}$, and the second one finally leaves 
us with $O(3,1)$ as residual symmetry group. As before, we will use 
latin indices again for the Lorentz quantities. The next step is 
the parameterization of a general state in terms of the Nambu-Goldstone bosons 
and the residual Higgs fields. We were not able to find a very elegant 
solution to 
this, but let us write down the following attempt:
\begin{equation}
g_{\alpha\beta} = r^a_{\ \alpha} r^b_{\ \beta} \tilde \eta_{ab}, 
\end{equation}
with 
\begin{equation}
\tilde \eta_{ab} = diag(1+\mu,\ -1+\nu,\ -1,-1). 
\end{equation}
The Nambu-Goldstone bosons  $r^a_{\ \alpha}$ correspond to the reducing matrix 
of the nonlinear realization approach and are, as such,  elements of 
$GL$. They also contain the parameterization of the (pure gauge) 
field $\phi$ (in their 
determinant). Since only those metrics can 
be related by a $GL$ transformation to the Minkowski metric $\eta_{ab}$, 
that possess the correct signature, in order to describe a general 
metric, we have to introduce two \textit{signature} functions $\mu$ and 
$\nu$. They are the residual Higgs fields and cannot be gauged away.
Clearly, they vanish in the groundstate. The form (66) is a modification 
of a similar parameterization proposed in \cite{19,28,29}. 

In practice, (66) is not really useful, in the sense that it does 
not allow us to express the Lagrangians (59) and (63), after 
gauging $r^a_{\ \alpha}$ away, in terms of $\mu$ and $\nu$ only, 
i.e., to separate the non-dynamical Minkowski metric from 
$\tilde \eta_{ab}$ and to carry out the variation with respect to 
$\mu $ and $\nu$ only. In view of this, nothing is really gained 
by the gauge transformation and we can equally well use 
 (59) and (63) in their initial form. Eventually, one has to come up 
with a better parameterization, that allows to express the metric 
directly in terms of the residual Higgs field, in a way that we 
explicitly see, for instance,  
the terms of the form $\mu_{,i}\mu^{,i}$ that are contained in the 
last term of (59).  

This, in principal, completes the analysis of the symmetry breaking 
mechanism of metric affine gauge theory. You can now go on and 
consider concrete models. The main features will be that, as 
long as nothing couples to the symmetric part of the connection (i.e.,, if 
there are no matter fields that possess a so called hypermomentum, see 
\cite{20}), the theory will remain in its groundstate and the 
signature will always be Lorentzian. Only fields with hypermomentum 
will provide the kinetic terms in (59) with a source and could be able 
to change the signature of $g_{\alpha\beta}$ and therefore, ultimately, 
of  $g_{ik}$. 

One can also consider the possibility of changing $\mathcal L_0$ 
by a factor, say, $(\det g_{\alpha\beta}) \phi^2$ (which is a scalar). 
This it actually preferable, since, then, you can express $\mathcal L_0$ 
without the help of the inverse metric $g^{\alpha\beta}$ and without 
dividing by $\det g_{\alpha\beta}$, making it suitable for the study 
of eventual signature changes. Similar as in the approach of 
Hehl et al., but with $g \phi^2$ playing the role of the scalar field, 
one is led to Brans-Dicke type theories. 
The analysis of
concrete models, however, is related to the description of matter fields 
with hypermomentum, which will in most cases include spinor representations 
of $GL$ (manifields), and is therefore
beyond the scope of this article.  

Let us also note that, in contrast to the approach in 
\cite{20}, the nonmetricity 
is not on the same level as the torsion or the curvature. It should 
appear explicitly only in the Higgs sector of the theory, as it does 
in the last term of (59). The Lagrangian 
$\mathcal L_0$ should, consequently,  be constructed from the torsion and 
the curvature tensors, which are the Yang-Mills tensors corresponding to the 
original gauge fields of the affine group. 
This does not prevent the nonmetricity (or rather the symmetric 
part of the connection, to which the nonmetricity reduces 
when  $g_{\alpha\beta}
= \eta_{\alpha\beta}$) to appear, 
even as a dynamical field, implicitly in terms like 
$R^{\alpha}_{\ \alpha ik}R^{\beta\ ik}_{\ \beta}$ .

Some critical remarks are at order considering the approach presented 
in this section. The scalar density $\phi$ was originally included 
to break the symmetry of $GL$ down to $SL^{\pm}$. This would be the 
case, if one were let to a nonzero groundstate described by 
\begin{equation}
|\phi(0)| = \alpha, 
\end{equation}
for some (nonzero) constant $\alpha$. 
Then, clearly, the residual symmetry is $SL^{\pm}$. However, (59) does 
not really do this job. Instead, it leads to (61), which does not exclude 
the value $\phi = 0$. Only if we assuming that $\phi \neq 0$
(in other words, if we assume that $GL$ is broken down to $SL^{\pm}$), 
we get the condition $g < 0$, which breaks the symmetry down to $O(3,1)$. 
Thus, the role of $\phi$ is actually to allow us 
to write down the Lagrangian (57) in a $GL$ invariant form and break 
the symmetry down to the Lorentz group, under the assumption that 
$\phi$ possesses a nonzero groundstate. It does not really determine that 
groundstate. The usual argument from $\phi^2 + \phi^4$
 Higgs theories, namely that the value $\phi = 0$ does not represent a 
minimum, but rather a maximum of the potential, and therefore is 
not a stable groundstate, is not valid in our 
case, because whether we are dealing with a maximum or a minimum 
depends directly on the sign of the metric determinant in (61), which is 
what we want to determine in the first place.    

In the dilaton approach in \cite{20}, 
the situation is quite different, because the 
dilaton field is used not to determine the signature of the metric, 
but merely to break the (additional) conformal symmetry of the theory. 

[There is actually a source of confusion in 
\cite{20}: The expressions local scale transformations, or dilations, 
refer to $GL$ transformations of the special form $G^{\alpha}_{\ \beta}
= \Omega \delta^{\alpha}_{\beta}$. We have adopted this vocabulary,
and consequently, call our field $\phi$ a \textit{dilaton} (since 
the effect of a general $G^{\alpha}_{\ \beta}$ on $\phi$ 
is identical to the effect of a dilation with  $\det G = \det \Omega$).  
In \cite{20}, 
however, the so-called dilaton field (section 6) is 
actually a $GL$-scalar (since 
it is invariant under the dilation $L=0, F=C=-1$, in the notation of 
\cite{20}), and the broken symmetry 
 is the conformal symmetry, not the 
scale, or dilation, invariance. More serious than this linguistic 
 problem is the fact that the authors, in the subsequent 
section (section 6.5), consider the dilational part of $GL$ as 
already broken by the (so-called) dilaton, and concentrate on the 
breakdown of $SL^{\pm}$ (or $T \ltimes SL$ in their notation). 
This is obviously mistaken, because even 
 if the (so-called) dilaton takes  a nonzero  
groundstate value, the broken symmetry will be the conformal one, 
leaving us therefore with the full, unbroken $GL$ symmetry group. 
Clearly, one cannot 
break the dilation invariance using a $GL$ invariant field.] 

This point needs further clarification. In contrast to the case of the 
translations, where the mere existence of a Poincar\'e (or affine) 
vector $y^a$ necessarily leads to a symmetry breaking, independently 
of the choice of the groundstate, in the case of the dilaton, the 
symmetry is only broken if the groundstate is different from zero. 
Otherwise stated, any vector $y^a$ is related by a translation 
to the groundstate $y^a = 0$, but only nonzero values for $\phi$ 
are related by a $GL$ transformation to the groundstate value, say, 
$|\phi| = 1$. 

One will thus have to come up with 
a potential that assigns a nonzero groundstate value to $\phi$. This 
can only be done if we have expressions where $\phi$ does not 
appear in the combination $\phi^2 \det g$, but this makes it 
rather difficult, if not impossible, 
 to construct  invariants. New ideas are needed here. 

For the moment, we will present a solution to the above problem, 
which  is based on the introduction of a scalar field $\psi$, 
in addition to the dilaton $\phi$. Consider the following 
Higgs Lagrangian:
\begin{equation}
\mathcal L_{higgs} = \sqrt{|g_{ik}|} \left [\psi(g\phi^2 + 1)+\psi(g^{ik}
\De_i g_{\alpha\beta} \De_k g^{\alpha\beta})\right],
\end{equation}
with $g = \det g_{\alpha\beta}$. 
The field equations for $\psi$ and $\phi$ read
\begin{eqnarray}
g\phi^2 + 1 + \De_i g_{\alpha\beta} \De^i g^{\alpha\beta} &=& 0\\
\phi g \psi &=& 0.
\end{eqnarray}
If the nonmetricity vanishes (which will be the case, again, if 
no matter fields couple to the symmetric part of the connection, i.e., 
if the kinetic term for $g_{\alpha\beta}$ has no source term), 
the first equation leads to the required groundstate characterization
\begin{equation}
g \phi^2 = -1.
\end{equation}
This excludes explicitely the values $\phi = 0$ and $g = 0$. Therefore, 
the symmetry, this time, has really been broken down to $O(3,1)$, without 
any ad hoc assumptions. Clearly, the metric determinant has to 
be negative, and therefore, the signature of $g_{\alpha\beta}$ is 
Minkowskian and we  can choose the groundstate $g_{\alpha\beta}(0)= 
\eta_{\alpha\beta}$. Any other choice, solution to (71), 
would be related by a $GL$
transformation
to this one and therefore be equivalent. From (71), you then also conclude 
that $|\phi| = 1$, and from (70) that $\psi = 0$. This last relation 
is rather interesting, since it means that in the groundstate, the 
whole Higgs sector is actually  zero, and therefore does not contribute 
to the Einstein and Cartan equations. In other words, as long as the 
nonmetricity is zero, the theory will be exactly equivalent to Poincar\'e 
gauge theory. Especially, there will be no contribution to the 
stress-energy tensor from the Higgs field. This has also the 
consequence that, as opposed to usual spontaneously broken theories, 
the gauge fields corresponding to the broken group generators will 
not acquire mass, since the term that is responsible for this, the 
last term in (68), vanishes together with $\psi$.

In (68), no kinetic terms for $\phi$ or $\psi$ have been included. 
This is also not 
necessary, since the actual Higgs field is $g_{\alpha\beta}$. 
The role of the field $\phi$ is merely to form a $GL$ invariant expression 
with the metric determinant, and 
 and $\psi$ is now the field that is introduced to break the 
equivalence of the 
tetrad and metric equations, i.e., to get an independent equation for the 
symmetry breaking mechanism. You may also see $\psi$ as a Lagrange 
multiplier. 

The Lagrangian (68), although it may appear artificial,  
is nevertheless the first Lagrangian  that \textit{really} does, what 
it is supposed to do, namely to break down the general linear group to 
the Lorentz group. It remains to see, if simpler solutions exist to 
this problem, that allow us to possibly eliminate one of the additional 
fields $\phi$ or $\psi$.  

On the other hand, one could also consider to start right from the beginning 
with a theory based on the 
 special affine group and avoid the problems we just described.  
This, however, seems rather unnatural 
and also breaks the direct link that exists between $GL$ theories 
and general relativity. Indeed, a strong argument in favor of $GL$ is 
the fact that we can gauge the tetrad into its trivial form 
$\delta^{\alpha}_i$, and then \textit{identify} the spacetime 
indices with the $GL$ indices. In order to maintain the gauge in 
the form $e^{\alpha}_{i} = \delta^{\alpha}_i$, it is necessary, every time   
 we change the coordinate system, to perform, at the same time, a $GL$ 
transformation with matrix $\partial \tilde x^i/\partial x^k $. 
 Under such a combined transformation, 
$\Gamma^{\alpha}_{\ \beta i}$ (then  written as $\Gamma^k_{\ li}$) 
will transform as a general relativity connection. Note that 
this also shows that, in a certain  way, the gauge approach presented in 
this article, contains as a subcase the spacetime based gauge 
approach of \cite{22} mentioned in the introduction. 
Indeed, fixing the gauge in the above way leads to a residual 
invariance in the form of the transformations considered in \cite{22}, 
and moreover,  tangent space $G/H$,  through $e^{\alpha}_i = 
\delta^{\alpha}_i$,  is identified with the spacetime manifold, an 
assumption that is made in \cite{22} and similar theories (e.g., \cite{13,14})
 right from the beginning. 

Also, the Lagrangian 
(63) will, in this gauge,  essentially reduce 
(up to torsion and nonmetricity parts 
contained in the connection) to the Einstein-Hilbert Lagrangian. 
This relation to general relativity is completely lost if we 
start with the special linear group. Note also, that theories 
based on the special affine group will lead to Brans-Dicke 
type theories, because we can gauge the metric $g_{\alpha\beta}$, 
even if it is in its groundstate signature,  
only up to a conformal factor into the Minkowski metric, leaving us 
therefore with a residual scalar Higgs field.  

Finally, and most importantly, we remind that the whole procedure, 
as described in this section, only works for the Lorentz stability 
subgroup. It is not possible, in a similar manner, to write down 
a Lagrangian that leads to a residual $O(4)$ or $O(2,2)$ symmetry, 
since in those cases, the mere knowledge of the sign of the metric 
determinant is not enough to determine its signature. In those 
cases, one would have to write down a Lagrangian in terms of 
a direct expression of the signature (which is a scalar), expressed 
in terms of the metric components. This may be possible, but it is 
not obvious to us, how this can be done in an simple way. Generally, 
the signature can be expressed in terms of the four 
eigenvalues $E_i$ of the metric, normalized to one, $e_i = E_i/|E_i|$. 
Those eigenvalues, however, are the  solutions  to a fourth order 
equation, and as such, highly nontrivial expressions in $g_{\alpha\beta}$. 

It seems therefore at least unlikely that a reasonable Lagrangian  
can be constructed 
that assigns a groundstate signature other then the Lorentzian one to 
the spacetime metric. 

Before we close this section, 
in order to avoid misunderstandings, let us clarify  
the critical remarks we made towards the references \cite{12,21} on 
one hand and \cite{20} on the other hand. We do not claim that there 
is anything wrong with the symmetry breaking mechanism based on 
the introduction of manifields, as proposed in \cite{12} and \cite{21}. 
It can be seen as an alternative way to trigger the symmetry breakdown 
from $SL$ down to the Lorentz group, and as such, should be completed 
by a suitable mechanism for the breakdown $GL \rightarrow SL$. 
What we claim is that, if one tries to perform the step $GL \rightarrow SL$ 
 with the help of a dilaton field in the way presented in 
this section,  
then the symmetry will break down to the Lorentz group anyway, 
and this makes the manifield superfluous. Nevertheless, the symmetry 
breaking mechanism described in \cite{12,21} is interesting by itself, 
and moreover, the infinite dimensional representations  of 
the special linear group described in those references  
will be needed anyway, if we want to complete 
our theory (in a $GL$-covariant way) with spinor fields. 
Thus, the manifields will enter, not as Higgs fields, but simply as 
matter fields, as generalization of the Dirac (and similar) Lagrangians. 
On the other hand, the (so-called) dilaton introduced in \cite{20} serves 
a different purpose, namely to break the conformal invariance  
of the theory's groundstate. Also here, the approach presents its 
own interests, in the context of theories with asymptotical conformal 
invariance at high energies, as described in \cite{20}. 
However, in addition to the conformal symmetry, we will also have to 
break the dilation invariance, i.e.,  include a mechanism for the 
$GL \rightarrow SL^{\pm}$ breakdown. 

Summarizing, in none of 
the above references, the symmetry breaking mechanism  
for the reduction  $GL\rightarrow O(3,1)$ is complete. 
Moreover, our approach is preferable in the sense that it 
clarifies the role of the metric tensor, 
and assigns to it the interpretation suggested by the nonlinear 
realization approach in \cite{19}. However, it remains to see if it 
is possible, either to avoid one of the additional fields, or, 
alternatively, to give a physical interpretation to those fields. 
As to the second possibility, one might consider $\phi$ and $\psi$   
as Higgs fields responsible for the breakdown of additional 
symmetries eventually present in $\mathcal L_0$. Possible candidates 
are, for instance, the projective symmetry, or, in the manner of Hehl 
et al., conformal invariance. 

In this article, we will consider yet another possibility, namely 
to enlarge the symmetry group to the special linear group $SL(R^5)$, in the 
hope 
to combine the (good) features of the de Sitter theory with the 
(seemingly incomplete) features of the affine theory.

\section{The group $SL(R^5)$}

In this section, $G^A_{\ B}$ ($A,B = 0,1,2,3,5$) 
denotes the elements of $SL(R^5)$ and greek indices denote the 
four dimensional part of $SL(R^5)$ quantities, i.e., $A = (\alpha,5)$ etc. 
The algebra of $SL(R^5)$ is given by the following relations:
\begin{equation}
[L^A_{\ B}, L^C_{\ D}] = i (\delta^A_D L^C_{\ B}-\delta^C_B L^A_{\ D}).
\end{equation}
The trace of the generators vanishes, $L^A_{\ A} = 0$, as a result 
of $\det G^A_{\ B} = 1$. In order to get a clearer picture about the 
contents of $SL(R^5)$, let us introduce the de Sitter metric 
$\eta_{AB} = diag(1,-1,-1,-1,-1)$. Then, the commutation relations 
can be written in the form 
\begin{eqnarray}
& [L_{\alpha\beta},L_{\gamma\delta}]= i(\eta_{\alpha\delta} L_{\gamma\beta} 
- \eta_{\gamma\beta} L_{\alpha\delta}),& \nonumber \\  
 & [P_{\alpha},L_{\gamma\delta}] = i(\eta_{\alpha\delta}P_{\gamma}),\ \  
 [K_{\beta}, L_{\gamma\delta}] = - i(\eta_{\gamma\beta}K_{\delta}), 
& \nonumber \\
& [P_{\alpha}, D] = - i P_{\alpha},\ \  [K_{\alpha}, D] = i K_{\alpha}, 
& \nonumber \\
& [P_{\alpha}, K_{\beta}] = -i(\eta_{\alpha\beta}D - L_{\alpha\beta}), 
& \nonumber \\ 
& [P_{\alpha}, P_{\beta}] = [K_{\alpha}, K_{\beta}]= [D, L_{\alpha\beta}] 
= 0, & 
\end{eqnarray}
with $P_{\alpha} = L_{\alpha 5},\ K_{\alpha} = L_{5\alpha},\ D = - L_{55}$. 
Replacing the 
$GL(R^4)$ generators $L_{\alpha\beta}$ with the Lorentz generators
 $L_{[\alpha\beta]}$, (73) takes 
the form of the algebra of the conformal group $SO(4,2)$. In view of 
this, we will refer to $P_{\alpha}, K_{\alpha}$ as 
generators of the pseudo-translations $G^{\alpha}_{\ 5}$, 
and the pseudo-inversions $G^5_{\ \alpha}$ (more precisely, pseudo-special 
conformal transformations), respectively.   

The group $SL(R^5)$ has also been considered in \cite{30} in the context 
of the nonlinear realization approach, 

Instead of the de Sitter  
metric, we introduce now a general Higgs metric $g_{AB}$, 
an $SL(R^5)$-tensor,  
which may be parameterized as follows 
\begin{equation}
g_{AB} = \left( \begin{array}{ccc} 
g_{\alpha\beta} + \phi^2 k_{\alpha} k_{\beta} 
&& \phi^2 k_{\alpha} \\ 
\phi^2 k_{\beta} && \phi^2 \end{array}\right).
\end{equation} 
The notation $g_{55} =\phi^2$ is used to be consistent with the 
notation of the previous section, as will become clear later. Although 
it turns out that $g_{55}$ is actually positive in the gauge we 
are interested in, and more generally, we are not interested in the 
global sign of the metric, 
for the moment,  there is no restriction on the sign of 
$\phi^2$. From (74), we find 
\begin{equation}
\det g_{AB} = \phi^2 \det g_{\alpha\beta},
\end{equation}
and, if the determinant does not vanish, the inverse metric reads 
\begin{equation}
g^{AB} = \left( \begin{array}{ccc} g^{\alpha\beta} && -k^{\alpha}\\
-k^{\beta} && k^2 + 1/\phi^2 \end{array}\right),
\end{equation}
where $g^{\alpha\beta}$ is the inverse of $g_{\alpha\beta}$ and 
$k^{\alpha} = g^{\alpha\beta}k_{\beta}, \ k^2 =k_{\alpha}k^{\alpha}$. 
In addition to $g_{AB}$, we introduce a further Higgs field, a
$SL(R^5)$-vector $y^A$, requiring the condition 
\begin{equation}
g_{AB}\ y^A y^B \neq 0. 
\end{equation}
With 20 Higgs fields $(g_{AB}, y^A)$, underlying one constraint, it 
should be possible to reduce the gauge group from the 24-parameter group 
$SL(R^5)$ to the 6 parameter Lorentz group. 

Similar to the de Sitter case (see equ. (29)), we introduce the 
spacetime metric 
\begin{equation}
g_{ik} = - \frac{v^2}{y^2}[ \frac{1}{y^2}(y_A \De_i y^A)(y_B \De_k y^B)
- \De_i y^A \De_k y^B g_{AB} ],
\end{equation}
where, as opposed to the de Sitter case, special care has to be given 
to the index positions ($y_A \De_i y^A \neq y^A \De_i y_A$). Note 
also that, in view of (29), the condition (77) actually has also been 
assumed (quietly) by Stelle and West, excluding thereby, for instance, 
the value $\phi = -v $ in (23). 

Let us also parameterize the $SL(R^5)$ connection as
\begin{equation}
\Gamma^A_{\ B} = 
\left( \begin{array}{cc} \Gamma^{\alpha}_{\ \beta} & \frac{1}{v} e^{\alpha} \\
B_{\beta} & A 
\end{array}\right).
\end{equation}
The constant $v$ is a length parameter that appears explicitly in the 
Higgs Lagrangian. Note that an $SL(R^5)$ invariant is given by the 
trace $\Gamma^A_{\ A}$. 
 
We propose the following Lagrangian 
\begin{equation}
\mathcal L_{higgs} = \sqrt{|g_{ik}|}\ [(y_A \De_i y^A)(y_B \De_k y^B) 
g^{ik} + \De_i g^{AB} \De_k g_{AB}\ g^{ik}- V(y,g)], 
\end{equation}
with the Higgs potential 
\begin{equation}
V(y,g) = (1+ \det g_{AB}) (g_{AB}y^A y^B - v^2) 
\end{equation}
Just as in the affine case, $V(y,g)$ is actually more an 
interaction of the different Higgs fields than a true potential. 
Recall that with the special linear group, the metric can be transformed 
into a non dynamical constant matrix,  up 
to a conformal factor. Therefore, the only truly dynamical quantity in 
$g_{AB}$, 
leading to an independent equation, is 
$\det g$, which is invariant under $SL(R^5)$. Thus, (81) leads to  
the groundstate characterization 
\begin{equation}
g_{AB} y^A y^B = v^2. 
\end{equation}
Before we analyze the symmetry breakdown in detail, we will try to make 
contact with the results of the previous section. In view of (82), 
we can choose the groundstate 
\begin{equation}
y^A(0) = (0,0,0,0,y^5(0)), 
\end{equation}
where the value $y^5(0)$ will depend on the groundstate value of $g_{55}$.
Let us choose $y^5(0) = v$ and $g_{55}(0)=1$.
 As in the de Sitter case, we can gauge a general state $y^A$ into the 
form 
\begin{equation}
y^A = (0,0,0,0,y^5).
\end{equation}
This leaves us with a residual gauge invariance 
\begin{equation}
G^A_{\ B} = \left( \begin{array}{cc} G^{\alpha}_{\ \beta} & 0 \\ q_{\beta} & s 
\end{array}\right), 
\end{equation}
 with the only restriction $s \det G^{\alpha}_{\ \beta} = 1$. Thus, 
$G^{\alpha}_{\ \beta}$ is a $GL(R^4)$ transformation. Let us not worry 
about the pseudo-inversions $q_{\alpha}$ for the moment. Consider the 
transformation 
\begin{equation}
G^A_{\ B} = \left( \begin{array}{cc} \delta^{\alpha}_{\beta} & 0 \\ 0 & s 
\end{array}\right). 
\end{equation}
This is not a $SL(R^5)$ transformation and has therefore to be accompanied 
by a $GL(R^4)$ transformation with inverse determinant, but nevertheless, 
it is instructive to see the transformation properties of the various 
fields under (86). For the metric components (74) we find 
\begin{equation}
\phi^2 \rightarrow \phi^2/s^2,\ \ g_{\alpha\beta}\rightarrow g_{\alpha\beta},
\ \ k_{\alpha} \rightarrow s k_{\alpha},
\end{equation}
and for the connection (79) 
\begin{equation}
\Gamma^{\alpha}_{\ \beta} \rightarrow \Gamma^{\alpha}_{\ \beta},\ \ 
e^{\alpha} \rightarrow  s^{-1}e^{\alpha}, \ \ 
B_{\alpha} \rightarrow s B_{\alpha},\ \ A \rightarrow A - s^{-1}\de s.
\end{equation}
Note also that $g_{ik}$ is an invariant under (86). This is a consequence 
of the fact that $g_{ik}$ is actually not only $SL(R^5)$, but also 
$GL(R^5)$ invariant. 
Let us recall the 
general classification of conformal transformations 
(in the metric affine framework) by Hehl et al. 
\cite{20}. Consider a function $\Omega$ and constants $C,L,F$. 
We call a \textit{projective} transformation 
the following 
\begin{equation}
\Gamma^{\alpha}_{\ \beta} \rightarrow \Gamma^{\alpha}_{\ \beta}- C
\delta^{\alpha}_{\beta} \ \de (\ln \Omega).
\end{equation}
By  \textit{(pure) conformal} transformation, we mean the following 
\begin{equation}
g_{\alpha\beta} \rightarrow \Omega^L g_{\alpha\beta},  
\end{equation}
and finally, a \textit{dilation} or \textit{scale transformation}
refers to a $GL(R^4)$ gauge transformation with 
\begin{equation}
G^{\alpha}_{\ \beta} = \Omega^F \delta^{\alpha}_{\beta}.
\end{equation}
Under the combined transformation (89)-(91), we find 
\begin{equation}
\Gamma^{\alpha}_{\ \beta} \rightarrow \Gamma^{\alpha}_{\ \beta}-(C+F)\ 
\delta^{\alpha}_{\beta}\ \de(\ln \Omega), \ \ \ 
g_{\alpha\beta} \rightarrow \Omega^{L-2F} g_{\alpha\beta},\ \ \   
e^{\alpha} \rightarrow \Omega^F e^{\alpha}. 
\end{equation}
Note that we have slightly changed the scheme of \cite{20}, in order 
to get a clearer separation of the different parts. The constant $C$ 
in \cite{20} (which we will denote by $C_{hehl})$ is related to our $C$ 
by $C_{hehl} = C + F$. The other constants are identical to those of 
Hehl et al. 

On the occasion, let us note that the so-called dilaton in \cite{20} 
transforms as $\sigma \rightarrow \Omega^{-L/2}\sigma$, i.e., it 
is invariant under dilations, as we have noticed in the previous section, 
and especially, it cannot be used to break the dilational invariance. 
On the other hand, our dilaton $\phi$ introduced in (59) 
transforms with the parameter $F$ and is therefore a true dilaton. 

Comparing (92) with (87) and (88), we see that the transformation (86) 
can be interpreted as a conformal transformation with parameters 
\begin{equation}
F = -1,\ \ C = 1,\ \ L = -2.  
\end{equation}
Especially, we see that a dilaton $\phi$ is automatically present in 
the theory. Indeed, the Higgs potential (81), in the gauge (84), using (75) 
 takes the form 
\begin{equation}
 V = \psi ( 1+ \phi^2 \det g_{\alpha\beta}), 
\end{equation}
with $\psi = g_{AB}y^Ay^B - v^2 = g_{55}y^5 y^5 -v^2$. In view of 
the groundstate $y^5(0) = v, g_{55}(0)= 1$, you can parameterize, 
for instance, a general state by  $y^5 = (1/\sqrt{g_{55}})\sqrt{v^2 + \psi}$, 
which leads to (94) immediately. The important thing is that, 
independently of the parameterization we choose, the factor $\psi$ is 
a scalar, whereas $\phi^2$ is a dilaton. Thus,  (94) is identical  
to  the Higgs  potential used in (68). 

The group $SL(R^5)$ therefore provides a natural explanation for 
the appearance of the dilaton and the scalar field which had to 
be introduced by hand in the framework of the affine group. 
In  contrast to (68), the Lagrangian (80) contains kinetic terms also 
for $\phi $ and $\psi$. This is quite natural, since now those fields 
appear as parts of the true Higgs fields $g_{AB}$ and $y^A$, 
while in (68), they played only a secondary role. 

Let us now return to (80) and (81) and complete the analysis of 
the symmetry breaking mechanism. As explained before, the 
only truly dynamical part of $g_{AB}$ is contained in its 
determinant. Therefore, under the groundstate conditions 
\begin{equation}
y_A \De_i y^A = 0,\ \ \De_i g_{AB} = 0, 
\end{equation}
the first of which is the direct generalization of the de Sitter case (34), 
we find the following equations
\begin{equation}
g_{AB}y^A y^B = v^2,\ \ y^A (\det g_{AB} + 1) = 0.
\end{equation}
In view of the first equation, there will certainly be a non-zero 
component of $y^A$, and therefore we find $\det g_{AB} = -1$. 
As before, we choose the groundstate 
\begin{equation}
y^A = (0,0,0,0,v).
\end{equation}
Then, the first equation in (96) leads to  $g_{55}(0) = 1$. 
Therefore, from $\det g_{AB}(0) = -1$ and with the help of (75), 
we are led to $\det g_{\alpha\beta}(0) = -1$, which means that 
we may choose the groundstate $g_{\alpha\beta}(0) = \eta_{\alpha\beta}$, 
the Minkowski metric with Lorentz signature. This leaves us with 
\begin{equation}
G^A_{\ B} = \left( \begin{array}{cc} 
\Lambda^{\alpha}_{\ \beta} & 0 \\ q_{\beta} & 1 
\end{array}\right), 
\end{equation}
as residual gauge invariance, with $\Lambda$ a Lorentz transformation. 
It is easy to show that the remaining parts of the metric, $g_{\alpha5}$, 
 can be gauged  away with a pseudo-inversion $q_{\beta}$. This leaves us 
with the groundstate 
\begin{equation}
g_{AB}(0) = diag(1,-1,-1,-1,1),\ \ y^A(0) = (0,0,0,0, v), 
\end{equation}
which is clearly Lorentz (and only Lorentz) invariant. 

As was the case with $GL(R^4)$, the parameterization of a general state 
is not trivial. Let us concentrate on those parts that cannot be 
transformed away. In other words, also in a general state, we 
can always fix the pseudo-translational and the pseudo-inversional gauge 
by requiring $y^{\alpha} = 0$ and $g_{\alpha 5} = 0$. Then, one 
might consider to use the $G^5_{\ 5}$ gauge to fix $g_{55} = 1$. 
Alternatively, one might fix $y^5$ to $v$ (not both, however). Let us 
take the first choice, and parameterize $y^5 = \sqrt{v^2 + \psi}$. 
At this point, the Higgs potential takes the form 
\begin{equation}
V = \psi (1 + \det g_{\alpha\beta}), 
\end{equation}
and the residual invariance group is $SL(R^4)$. 
Note also that (78) takes the expected form 
$g_{ik} = e^{\alpha}_i e^{\beta}_k g_{\alpha\beta}$.
The dilaton seems to 
have disappeared, but remember that we can transform $g_{\alpha\beta}$ 
only up to a conformal factor into a constant matrix. Thus, for 
the four dimensional part of the metric, using 
a similar parameterization as in (65), (66), and fixing the gauge 
by transforming away the Nambu-Goldstone bosons, we can write   
\begin{equation}
g_{ab} = \phi^2 \tilde \eta_{ab}, 
\end{equation}
with 
\begin{equation}
\tilde \eta_{ab} = diag(1+\mu,\ -1+\nu,\ -1,-1). 
\end{equation}
We use latin indices again, to underline the residual Lorentz invariance. 
Again, we can consider $\phi^2$ to be positive, since the global sign 
of the metric is not important. 

Thus, the broken, Lorentz invariant Lagrangian contains 4 residual 
Higgs fields, $\phi, \mu, \nu$ and $\psi$. At this stage, they 
transform simply as scalar fields. Their kinetic terms are contained in 
(80), but again, it will be difficult to find an explicit form in 
terms of $\mu$ and $\nu$ and it might be preferable go one step back 
and use an $SL(R^4)$ invariant form.   

The full analysis of the $SL(R^5)$ theory is beyond the scope of 
this article. The theory, compared to  the metric affine theory,  
contains five additional fields, $B_a$ and $A$ which will have to 
be interpreted and whose coupling to matter fields will have to be 
discussed. The incorporation of spinor fields in our theory can 
be done using the $SL$ spinor representations of Ne'eman and 
Sijacki \cite{12,21}. 
As opposed to the metric affine theory, where this is only 
possible after reducing $GL(R^4)$ down to $SL(R^4)$,  
a fully $SL(R^5)$-invariant Lagrangian with spinor matter 
can thus be written down.

\section{Nambu-Goldstone 
fermions and $GL(R^4)$ invariant Dirac Lagrangian}

Finally, we will briefly sketch a quite different model of a 
dynamically broken gravitational theory. Although the theory, as it  
will be presented here, is still incomplete, especially with respect to   
the interpretation of the tetrad field and the incorporation of translations, 
it is nevertheless rather promising under several aspects. First, 
it enables us to write down  a $GL(R^4)$ covariant equation 
for a simple vector field $k^{\alpha}$ (as opposed to an infinite 
dimensional $SL(R^4)$ spinor),  that, 
after the symmetry breaking, reduces to the usual form of the  
Dirac equation without any ad hoc assumptions. Moreover, 
the fermion fields appear under a very different and rather 
instructive light, compared to the usual derivation of the Dirac 
equation as \textit{square root} of the Klein-Gordon equation. 

Let us go back to the nonlinear realization approach of section 
2. Apart from the connection (5), we have claimed that 
we can, for every field $\phi$ transforming under \textit{some}
representation of $G$, define the field $\psi = \sigma^{-1} \phi$, that 
will transform under \textit{some} representation of $H$. It is usually 
understood that, if $\phi$ transforms under a vector (spinor) representation 
of $G$, then $\psi $ too will transform under a vector (spinor) representation 
of $H$. This means that $\sigma$ itself will sometimes be interpreted as 
a mixed $G-H$ tensor, and sometimes as a mixed $G-H$ spinor. This is the 
usual conception in the nonlinear realization approach. 

This may make sense, as long as one considers $\sigma$ simply as 
transformation that parameterizes the degrees of freedom of the 
coset space $G/H$. However, as we have seen, in a dynamically broken 
theory, $\sigma$, represented by the reducing matrix $r^{\alpha}_{\ a}$, 
 will appear as the Nambu-Goldstone field of the theory. These fields, 
even though of pure gauge nature, describe the Nambu-Goldstone particles, 
and as such, have to transform under a well specified representation 
of $G$ and of $H$. We have thus to decide whether $r^a_{\ \alpha}$ 
transforms as $H-G$ tensor (as was the case throughout our article) 
or, say, as $H-G$ spinor. 

In this section, we are interested in the breakdown of the group 
$GL(R^4)$ to the Lorentz group. Since spinor  representations of 
 $GL(R^4)$ are necessarily infinite dimensional, we choose here 
a third way. We introduce the reducing matrix with the 
following transformation behavior (see (9)):
\begin{equation}
r^M_{\ \alpha} \rightarrow L^M_{\ N} r^N_{\ \beta}(G^{-1})^{\beta}_{\ \alpha},
\end{equation}
with $G^{\alpha}_{\ \beta} \in GL(R^4)$ and  
\begin{equation}
L^M_{\ N} = e^{i \epsilon^{ab} (\sigma_{ab})^M_N}, 
\end{equation}
where $\sigma_{ab} = (i/2) [\gamma_a, \gamma_b]$ are (twice) the generators 
of the spinor representation of the Lorentz group and 
$\epsilon_{ab} \in R$. In other words, $r^M_{\ \alpha}$ transforms 
as a spinor with respect to the index $M$. More generally, 
from a $GL(R^4)$ vector $k^{\alpha}$, we can now form 
the quantity 
\begin{equation}
\psi^M = r^M_{\ \alpha} k^{\alpha}, 
\end{equation} 
which transforms exactly like a Dirac spinor, $\psi \rightarrow 
 e^{i\epsilon^{ab}\sigma_{ab}}\psi$. Usually, this is called an 
$SL(2,C)$ transformation, and we will follow that tradition,  
although whenever we talk about 
spinors, we actually refer to 4-component Dirac spinors, 
or bispinors. Note also that we use capital latin letters from the 
middle of the alphabet to index the $SL(2,C)$ quantities, 
as opposed to $a,b,c,\dots$ which will be used, as before, for 
$SO(3,1)$ indices.  

We see that $r^M_{\ \alpha}$ is very similar to a spin 3/2 particle 
$\psi_i$, or $\psi^M_i$ if we write the Dirac space index explicitely,   
 with the only difference that the spacetime index $i$ is replaced by 
a $GL(R^4)$ index. As we know, we can always choose a gauge in 
which we can identify those indices. Therefore, the corresponding 
Nambu-Goldstone particles will be of fermion nature.

An  $SL(2,C)$ connection can also be defined using (8) and $r^M_{\ \alpha}$. 
More physical than \textit{reducing} the symmetry by hand 
is, of course, the dynamical Higgs approach, to which we 
will turn now. 

As gravitational fields, we introduce a $GL(R^4)$ connection 
$\Gamma^{\alpha}_{\ \beta}$ and a tensor valued one-form 
$e^{\alpha}_{\ \beta}$, whose relation to the tetrad field will 
be clarified in the following. In addition, instead of the 
metric $g_{\alpha\beta}$ of the previous sections, we consider 
a Higgs field $\gamma^{\alpha}_{\ \beta}$ with one co- and one 
contravariant index. 

Before we construct the gravitational Lagrangian, let us point out 
some important differences between the metric $g$ and the tensor $\gamma$. 
In contrast to the metric tensor, which  
transforms as $g \rightarrow G^{-1} g (G^{-1})^T$, the tensor $\gamma$  
will transform as $\gamma \rightarrow G \gamma G^{-1}$. Therefore, 
while the metric is entirely characterized by its signature, $\gamma$ 
will be characterized by its four eigenvalues. These cannot be 
scaled to $\pm 1$. More generally, if we assign a groundstate to $\gamma$, 
there will always be a residual dilational invariance $G^{\alpha}_{\ \beta} 
= \lambda \delta^{\alpha}_{\ \beta}$ under which $\gamma$ is invariant. 
On the other hand, in contrast to $g$, from $\gamma$ we can form  
the  $GL(R^4)$ invariants $\det \gamma$ 
and $Tr\  \gamma$. 

Since $\gamma$ contains a priori 16 independent components, while 
the broken gauge degrees of freedom are only 10, we are free to 
put constraints on $\gamma$. It turns out that it is enough to 
require 
\begin{equation}
\gamma^{\alpha}_{\ \alpha} = 0.
\end{equation}
Let us define the gauge invariant spacetime metric as 
\begin{equation}
g_{ik} = - e^{\alpha}_{\ \beta i} e^{\beta}_{\ \alpha k}. 
\end{equation}
Consider the following Higgs potential: 
\begin{equation}
\sqrt{|g_{ik}|}\  V = \sqrt{|g_{ik}|}\  
[ \frac{1}{4} Tr (\gamma^4) + \frac{1}{2} Tr (\gamma^2)]. 
\end{equation} 
The matrix notation is to be understood as $Tr (\gamma^2) 
= \gamma^{\alpha}_{\  \beta} \gamma^{\beta}_{\ \alpha}$ and 
similar for $Tr (\gamma^4)$. Variation with respect to $\gamma$ leads 
to the groundstate characterization 
\begin{equation}
\gamma^{\alpha}_{\ \beta} \gamma^{\beta}_{\ \delta} 
= - \delta^{\alpha}_{\  \delta}.
\end{equation}
The other solution, $\gamma = 0$, corresponds to a maximum of $V$ and 
is thus unstable.  
Although we did not explicitely mention it, it is understood that 
all the fields are real, since we are dealing with a $GL(R^4)$ theory. 

The condition (109), together with the constraint (106), are enough 
to characterize $\gamma$ completely. We see this as follows: Suppose 
we diagonalize $\gamma$. Then, (106) and (109) lead to the 
eigenvalues $(i,i,-i,-i)$. In other words, the eigenvalues are entirely fixed,
and therfore $\gamma$ itself is fixed up to similarity transformations.  
In our specific case, we cannot diagonalize the real matrix $\gamma$ with a 
$GL(R^4)$ transformation, but this is also not necessary. We can choose 
any real representative of $\gamma$ (with eigenvalues $(i,i,-i,-i)$), say 
\begin{equation}
\gamma = \left( \begin{array}{cccc} 0&1&0&0\\-1&0&0&0\\0&0&0&-1\\0&0&1&0 
\end{array}\right) = i \left( \begin{array}{cc} \sigma^2 & 0\\ 0 
& - \sigma^2 \end{array}\right)
\end{equation}
and then, any other real solution to (106) and (109) will be related 
to this choice by a $GL(R^4)$ transformation.  

The groundstate (110) certainly breaks the $GL(R^4)$ symmetry. But down
to which subgroup? As mentioned earlier, the dilational symmetry 
cannot be broken. In order to find the rest of the  residual symmetry group, 
it is convenient to introduce, at this point,  the  following 
Dirac matrices 
\begin{eqnarray}
\gamma_0 = 
\left( \begin{array}{cc} 
0 & \sigma^2 \\ \sigma^2 & 0 
\end{array}\right), \ \ \  
& \gamma_1 =  
\left( \begin{array}{cc} 
-i \sigma^3  & 0 \\ 0 & -i \sigma^3 
\end{array}\right), \  \ \ 
\nonumber \\
\gamma_2 = 
\left( \begin{array}{cc} 
0 & \sigma^2\\ -\sigma^2 & 0
\end{array}\right), \ \ \   
& \gamma_3 = 
\left( \begin{array}{cc} 
 i \sigma^1 & 0 \\ 0 &i \sigma^1
\end{array}\right). 
\end{eqnarray} 
These are the usual Dirac matrices in the Majorana representation. 
We have the following relation: 
\begin{equation}
\gamma_5 \equiv \gamma = \gamma_0 \gamma_1 \gamma_2 \gamma_3. 
\end{equation}
We attach the index $5$ to our groundstate Higgs matrix, in order to 
indicate its relation to the Dirac matrices in the specific 
representation (and not to denote any kind of transformation 
behavior). Note that our $\gamma_5$ differs by a factor $i$ from 
the usual definition. 

Consider the exponential map, i.e., write the  $GL(R^4)$ 
transformation (in the neighborhood of the identity $I$) in the form 
\begin{equation}
G^{\alpha}_{\ \beta} = e^{\sigma^{\alpha}_{\ \beta}}
\end{equation} 
where $\sigma$ some real $4 \times 4$ matrix. [Similarly, the elements 
in the neighborhood of $-I$ can be written as 
$-\exp{(\sigma^{\alpha}_{\ \beta})}$. It is, however, evident 
that, if (110) is invariant under some transformation $G$, then also 
under $-G$.] 
As to the other connected 
component of $GL(R^4)$ (with $\det G < 0$), its elements can be 
written in 
the form 
\begin{displaymath}
G^{\alpha}_{\ \beta} = \pm P e^{ \sigma^{\alpha}_{\ \beta}}, 
\end{displaymath}
with 
\begin{displaymath} 
P = \left( \begin{array}{cccc}1&0&0&0\\0&-1&0&0\\0&0&-1&0\\0&0&0&-1
\end{array}\right),
\end{displaymath} 
the spatial inversion (or parity) transformation. It is easily 
checked that (110) is not invariant 
under $P$, and therefore, we can concentrate on the transformations 
of the form (113). 
The invariance of $\gamma^{\alpha}_{\ \beta}$, i.e., the 
requirement $G \gamma_5 G^{-1} = \gamma_5$, leads 
infinitesimally to the following condition 
\begin{equation}
\sigma \gamma_5 = \gamma_5 \sigma.  
\end{equation}
It is now an easy task to determine the matrices $\sigma$ that satisfy 
this condition. It is convenient to consider, temporarily, $\sigma$ 
to be a complex matrix.  
We know from Dirac theory that the complete set of  
 linearly 
independent (complex)  matrices is  given by the sixteen matrices 
$(I, \gamma_5, \gamma_a, \gamma_5 \gamma_a, [\gamma_a, \gamma_b])$.  
Apart from $I$ and $\gamma_5$, corresponding to dilations and 
chiral transformations, to which we will turn later on, the only matrices 
satisfying (114) are the six matrices $[\gamma_a, \gamma_b]$. Therefore, 
the most general transformation allowed by (114) (up to dilations and chiral 
transformations), has 
the form 
\begin{equation}
G = e^{\tilde \epsilon^{ab} [\gamma_a, \gamma_b]}, 
\end{equation}
with, in general, complex parameters $\tilde\epsilon^{ab}$. Let us 
introduce the Lorentz generators 
\begin{equation}
\sigma_{ab} = \frac{i}{2} [\gamma_a, \gamma_b]. 
\end{equation}
More precisely, the matrices $\sigma_{ab}/2$ are the Lorentz generators, 
but it is customary to define $\sigma_{ab}$ in the form (116).  
We see from (111) that all six $\sigma_{ab}$ are purely imaginary. 
Therefore, 
since the exponent in (115) is real,  we conclude 
that $G$ has the form 
\begin{equation}
G = e^{-\frac{i}{4} \epsilon^{ab}\sigma_{ab}},
\end{equation}
with real parameters  $\epsilon^{ab}$. (The factor $-1/4$ is 
conventional.)
This, however, is exactly the $SL(2,C)$ transformation (104). 
Note that the essential step was 
the reality requirement of the $GL(R^4)$ transformation. Without 
that, we can, in (115), replace any matrix $\gamma_a$ by $i \gamma_a$, 
which is equivalent to a signature change of the metric 
$\eta_{ab}= \frac{1}{2} (\gamma_a \gamma_b+ \gamma_b \gamma_a)$ and 
more generally to a change of the residual symmetry group. The reality 
requirement, on the other hand, allows only for a change $\gamma_a \rightarrow
 i\gamma_a$ of \textit{all} matrices, corresponding to a global, irrelevant, 
sign change of the metric.   

As before, we use the letters  $M,N\dots$ to denote $SL(2,C)$ quantities. 
The requirement that the Dirac matrices $\gamma^a = (\gamma^a)^M_N$  
are invariant under a gauge transformation leads to the usual relation 
between Lorentz and $SL(2,C)$ transformations, i.e., 
\begin{equation}
\gamma^a 
\rightarrow e^{-\frac{i}{4}\epsilon^{cd} \sigma_{cd}} 
(\Lambda^a_{\ b}\gamma^b)
e^{\frac{i}{4} \epsilon^{cd}\sigma_{cd}} = \gamma^a,  
\end{equation}
which is satisfied (infinitesimally) for 
$\Lambda^a_{\ b} = \delta^a_b + \epsilon^a_{\ b}$. [The same 
can be achieved replacing $G$ by $-G$ in (117). This reflects  the 
well known two-to-one correspondence between $SL(2,C)$ and $SO(3,1)$.] 
After the symmetry 
breakdown, the $GL(R^4)$ connection $\Gamma^{\alpha}_{\ \beta}$ reduces 
to a $SL(2,C)$ connection $\Gamma^M_{\ N}$ and the $GL(R^4)$ tensor 
$e^{\alpha}_{\ \beta}$ to the $SL(2,C)$ tensor $e^M_{\ N}$. Following 
Chamseddine \cite{31}, we can  
introduce $SO(3,1)$ components  in the following 
way
\begin{eqnarray}
e^M_{\ N} &=& i e^a (\gamma_a)^M_N, \nonumber \\
\Gamma^M_{\ N} &=& - \frac{i}{4} \Gamma^{ab}(\sigma_{ab})^M_N.
\end{eqnarray}
Since   $\Gamma^M_{\ N}$ and $e^M_{\ N}$ are real, the same holds for  
the Lorentz connection $\Gamma^{ab}$ and the tetrad field $e^a$. 
The inverse relations are 
\begin{eqnarray}
e^a &=& -\frac{i}{4}\ Tr\ ( e\ \gamma^a), \nonumber \\
\Gamma^{[ab]} &=& \frac{i}{2}\ Tr\ (\Gamma\ \sigma^{ab}), 
\end{eqnarray}
where the traces are to be taken in Dirac space, and $e = e^M_{\ N},\  
\Gamma = \Gamma^M_{\ N}$. 
We see that (119) parameterizes only the antisymmetric part of the 
connection $\Gamma^{ab}$. The non-metricity parts are lost. We 
will return to this point later on. 
Note also that the spacetime metric (107) takes the expected form 
$g_{ik} = e^a_i e^b_k \eta_{ab}$. 

Let us now construct the Lagrangian of the theory. Consider the 
following $GL(R^4)$ invariant Lagrangian: 
\begin{equation}
\mathcal L_0 = \frac{1}{4} 
\gamma^{\alpha}_{\ \beta} R^{\beta}_{\ \gamma} \wedge 
e^{\gamma}_{\ \delta}\wedge e^{\delta}_{\ \alpha}. 
\end{equation}
In the groundstate $\gamma = \gamma_5$, this simply reduces to 
\begin{equation} 
\mathcal L_0 = \frac{1}{4} Tr [ \gamma_5 \ R \wedge e \wedge e ], 
\end{equation}
where we use the matrix notation, as is customary in Dirac space. 
This, however, is exactly Chamseddine's form of the  $SL(2,C)$ invariant 
Einstein-Cartan Lagrangian \cite{31}. Indeed, using (119) and (120), we 
find $R^M_{\ N} = -(i/2)R^{ab}(\sigma_{ab})^M_N$ and  
\begin{equation}
\mathcal L_0 = \frac{1}{2} \epsilon_{abcd} R^{ab} \wedge e^c \wedge e^d, 
\end{equation}
or, in the conventional form, $\mathcal L_0 =  -2 e R$.  
 
Until this point, nothing seems to have been gained compared to the 
usual metric approach of the affine theory. On the contrary, several 
issues are still open,  
like the discussion of the dilations and chiral transformations, 
as well as the parameterization of 
the non-metricity. We postpone the full analysis of 
those problems to future work, although 
some insight will also be gained in the following. 

However, when it comes to spinor fields, the approach of this section 
clearly presents undeniable advantages. This is the point we concentrate 
on in this article. 

The presence of the tensor valued one-form $e^{\alpha}_{\ \beta}$ 
allows us to write down a Lagrangian for a vector field $k^{\alpha}$ 
of first order in the derivatives. More precisely, consider 
the two vector fields $k^{\alpha}$ and $k_{\alpha}$. (Vectors under 
$GL(R^4)$, scalars under spacetime transformations.)
Since we are 
not in the possession of a $GL(R^4)$ metric that could relate a 
covariant to a contravariant vector, these fields have to 
be considered as independent from each other. 
Moreover, we allow both fields to take 
complex values, with the only restriction $k_{\alpha} k^{\alpha} \in R$. 

Consider the Lagrangian 
\begin{eqnarray}
 \mathcal L_M 
&=& 
\frac{1}{48}\left[ k_{\alpha} e^{\alpha}_{\ \beta}\wedge e^{\beta}_{\ \gamma}
\wedge e^{\gamma}_{\ \delta}\wedge 
\gamma^{\delta}_{\ \epsilon} \De k^{\epsilon} 
 -  
\De k_{\alpha}\wedge  e^{\alpha}_{\ \beta}\wedge e^{\beta}_{\ \gamma}
\wedge e^{\gamma}_{\ \delta} 
\gamma^{\delta}_{\ \epsilon}  k^{\epsilon}\right] \nonumber \\ &&
 + \sqrt{|\det g|} \ m  k_{\gamma}k^{\gamma}, 
\end{eqnarray} 
with $\De k_{\alpha} = \de k_{\alpha} - \Gamma^{\beta}_{\ \alpha}k_{\beta}$
and $\De k^{\alpha} = \de k^{\alpha} + \Gamma^{\alpha}_{\ \beta}k^{\beta}$. 
We claim that (124) is the Dirac Lagrangian in a general linear invariant 
form. Indeed, in the groundstate $\gamma = \gamma_5$, using the 
parameterizations (119) and changing, as in (105),  
the notation to  $k^M = \psi^M$ 
and $k_M = \bar \psi_M$,  (124) reduces to  
\begin{eqnarray} 
 \mathcal L_M &=& -\frac{i}{12}\left[ 
\bar \psi (e^a \wedge e^b \wedge e^c \epsilon_{abcd} 
\gamma^d) \wedge \De \psi  
- \frac{i}{12}\  \De \bar \psi \wedge
(e^a \wedge e^b \wedge e^c \epsilon_{abcd} 
\gamma^d)  \psi \right] \nonumber \\ && +  e\  m \bar\psi \psi \nonumber \\
&=& e\left(\frac{i}{2} \left[\bar \psi  \gamma^m \De_m \psi 
 - (\De_m \bar \psi) \gamma^m \psi \right]+  m \bar \psi \psi\right)
\end{eqnarray}
with $\De \psi = \de \psi - \frac{i}{4} \Gamma^{ab}\sigma_{ab}\psi$, 
$\De \bar \psi = \de \bar \psi + \frac{i}{4} \bar 
\psi \Gamma^{ab}\sigma_{ab}$, $\gamma^m = e^m_a \gamma^a$ and 
$e = \det e^a_m$. This is   
 the well known form of 
the Lorentz invariant Dirac Lagrangian in the framework of Poincar\'e gauge
theory \cite{20}.   

The complete Lagrangian will be composed from (108), (121) and (124), 
and possibly a kinetic term for  the Higgs field $\gamma$. 
  
It is easy to check that $\psi, \bar \psi, \De \psi$ and $\De \bar \psi$ have 
the required $SL(2,C)$ transformation behavior known from Dirac theory. 
This is evident anyway, by construction. There remains one point to 
discuss, namely the relation between $k^{\alpha}$ and $k_{\alpha}$, or 
$\psi$ and $\bar \psi$ respectively. 

We have argued that $k_{\alpha}$ and $k^{\alpha}$ have to be 
considered as independent, due to the lack of a $GL(R^4)$ metric. 
However, after the symmetry breakdown, the residual symmetry 
is $SL(2,C)$ (neglecting again the dilations and chiral transformations). 
Similar to 
the Minkowski metric in the Lorentz case, one might look for a 
non-dynamical, constant  tensor that is invariant under $SL(2,C)$. 
In other words, is there an invariant tensor $g_{MN}$ that could 
relate covariant and contravariant $SL(2,C)$ vectors in the form 
(take care of the index positions, $g_{NM}$ may by asymmetric)
\begin{equation}
k_M = g_{NM} k^N? 
\end{equation}
Note that, as opposed to the $\gamma^a$ matrices, $g_{MN}$ has two 
lower indices and therefore transforms as
\begin{equation}
g \rightarrow e^{-i \epsilon^{ab} 
\sigma_{ab}} g (e^{-i \epsilon^{ab} \sigma_{ab}})^T. 
\end{equation}
Infinitesimally, the requirement $g\rightarrow g$ takes the form 
\begin{equation}
\sigma g + g \sigma^T = 0, 
\end{equation}
with   $\sigma $ one of the matrices $\sigma_{ab}$. Considering again 
the 16 linearly independent 4x4 matrices, we find that there is 
only one matrix that anti-commutes in this way with all the matrices 
$\sigma_{ab}$, namely 
\begin{equation}
g_{MN} = \alpha (\gamma_0)_{MN}.
\end{equation}
Note, by the way, that $\gamma_5$ does not anti-commute in this way with 
$g_{MN}= (\gamma_0)_{MN}$, and therefore, the requirement of the 
existence of the $SL(2,C)$ invariant metric also breaks the 
chiral invariance. Finally, the requirement that $g_{MN}$ is 
a non-dynamical, constant matrix (i.e., $\alpha = const$),  
also breaks the dilational 
gauge freedom. We may choose $\alpha = 1$. Let us emphasize that the 
matrix $(\gamma_0)_{MN}$ in (129) has nothing to do with the Dirac matrix 
$(\gamma_0)^M_N$. Apart from the different index positions, there is 
also the important difference that in (129), the index $0$ does not mean 
that the matrix is a component of a Lorentz vector (as is the case with 
$(\gamma_0)^M_N$), but it is attached only to indicate the (rather 
incidental) numerical coincidence of both matrices.  

However, the relation $k_M = (\gamma_0)_{NM} k^N $ is easily shown 
to lead to $k_M k^M = 0$ for any $SL(2,C)$ vector $k^M$. 
Therefore, we replace 
(126) by 
\begin{equation}
k_M = g_{NM} (k^N)^*,\ \  \text{with} \ \ g_{NM} = (\gamma_0)_{NM},   
\end{equation}
 where the star denotes complex conjugation. Note that, for real $k^M$, 
we are lead to $k^M k_M = 0$. This is the reason why it is necessary 
to introduce complex fields. [Recall in this context that, in the 
Majorana representation, 
the charge conjugation operator $C$ is given by $- \gamma^0$ and 
therefore $\psi_c = \psi^*$.] 
In matrix notation, and 
writing  again $k^M = \psi^M, k_M = \bar \psi_M $, (130) reads 
\begin{equation}
\bar \psi = \psi^t \gamma_0, 
\end{equation}
with $\psi^t = (\psi^*)^T$ the hermitian conjugate. This is the well 
known relation for the adjoint spinor in Dirac theory. Note that 
(130) assures that $k_M k^M$, i.e., $\bar \psi \psi$,  
is automatically real, which justifies 
the choice of $\alpha = 1$ in (129).  

We conclude that the requirement of the existence of an $SL(2,C)$ 
invariant matrix, that plays the role of a metric tensor in 
Dirac space, in the sense of equation (130), leads to the usual 
relation between the spinor and the adjoint spinor field. The 
introduction of this tensor, which is numerically equal to $\gamma_0$,  
also breaks the dilational and chiral invariance 
that were still symmetries of the groundstate  (110). 

The  metric (129) is also useful under another aspect. It can be used 
to raise the second index of the $SL(2,C)$ connection $\Gamma^M_{\ N}$, 
which allows us to consider the symmetric and antisymmetric parts of 
$\Gamma^{MN}$. It is then easy to show that the antisymmetric part 
transforms as tensor under $SL(2,C)$, whereas the symmetric part 
has a non-homogeneous transformation behavior. The situation is 
thus reversed  compared to  the Lorentz connection $\Gamma^{ab}$. 
It also becomes clear that the parameterization (119) presupposes 
that $\Gamma^{MN}$ and $e^{MN}$ are symmetric, because the matrices 
$\sigma_{MN}$ and $(\gamma^a)_{MN}$ (where one index has been lowered 
with $g_{MN} = (\gamma_0)_{MN}$) are symmetric. 

Therefore, in order to parameterize completely the 
connection $\Gamma^M_{\ N}$, one will  have to use the sixteen linearly 
independent matrices in Dirac space and write, instead of (119),  
\begin{equation}
\Gamma^M_{\ N} = - \frac{i}{4}\ \Gamma^{ab} (\sigma_{ab})^M_N 
+ i A^a (\gamma_a)^M_N + iB^a (\gamma_a \gamma_5)^M_N 
+  C \delta^M_N 
+ D (\gamma_5)^M_N, 
\end{equation}
and in the following, analyze the transformation behavior 
of the one-forms $A^a, B^a, C, D$, as well as their coupling 
to the Dirac field. This parameterization of the $GL(R^4)$ connection
 in terms of the irreducible components under the Lorentz group is 
rather similar to the case of the $SO(4,2)$ connection (conformal group) 
(see \cite{16} for instance), the difference lying in the 
dilational part $C \delta^M_N$ which is not present in the conformal case. 
One will also have to put certain constraints on 
$e^{\alpha}_{\ \beta}$,  because else, $e^M_{\ N}$ will allow for a similar 
parameterization, and the direct correspondence to $e^a$ will 
be lost. 

In our quest for a minimalistic model, there is also the attractive 
idea to consider the tetrad as being related to the (pseudo)vector part 
$A^a$ or $B^a$
of the connection (132). Indeed, with (132), the curvature tensor 
(after the symmetry breaking) takes the following form 
\begin{eqnarray}
 R &=& \de \Gamma + \Gamma \wedge \Gamma \nonumber \\
 &=& - \frac{i}{4} R^{ab} \sigma_{ab} +i(A^a\wedge A^b + B^a \wedge B^b)
\sigma_{ab} 
+ i(\de A^a + \Gamma^a_{\ c}\wedge A^c) \gamma_a \nonumber \\ && - 2 i
(B^a \wedge D)\gamma_a 
 + i (\de B^a + \Gamma^a_{\ c}\wedge B^c) \gamma_a \gamma_5 + 2 i
( A^a \wedge D)\gamma_a \gamma_5 \nonumber \\ && 
 + \de D \gamma_5 - 2(A^a \wedge B_a) \gamma_5 + \de C. 
\end{eqnarray}
Therefore, one could consider, instead of (121) the following Lagrangian 
\begin{eqnarray}
 \mathcal L_0 &=& Tr\ (\gamma R \wedge R) \nonumber \\
& =& - 4 (\de D - 2 A^a \wedge B^a) \wedge \de C  \nonumber \\ 
&& - 4 \epsilon_{abcd} 
[\frac{1}{4} R^{ab}- A^a \wedge A^b - B^a \wedge B^b]   
\nonumber \\ && \ \ \ \ \ \ \ \ \wedge 
[\frac{1}{4} R^{cd}- A^c \wedge A^d - B^c \wedge B^d].   
\end{eqnarray}
The first line is the $GL(R^4)$ invariant form, depending only on 
$\Gamma^{\alpha}_{\ \beta}$ and the Higgs field $\gamma^{\alpha}_{\ \beta}$, 
and the following lines are its decomposition in the groundstate $\gamma = 
\gamma_5$.  
Under the assumption that the pseudo-vector part $B^a $ vanishes, 
we can identify $A^a \sim e^a$, and 
(134) reduces to the Einstein-Cartan Lagrangian with cosmological constant. 
(The terms $\de C \wedge \de D$ and $\epsilon_{abcd} 
R^{ab} \wedge R^{cd}$ are 
total divergences.)

In order to write down the Dirac Lagrangian (124), we define 
the tensor valued one-form 
\begin{equation}
e \equiv  \frac{1}{2} \gamma \De \gamma = \frac{1}{2} \gamma \de \gamma +
\frac{1}{2} \gamma [\Gamma, \gamma], 
\end{equation}
which decomposes in the groundstate to 
\begin{equation}
e = iA^a \gamma_a  + i B^a \gamma_a \gamma_5.
\end{equation}
Again under the assumption $B^a = 0$, this reduces to the 
the form of $e$ given in (119), and can be used in (124), as well as 
in (107),(108). Of course, one will 
have to justify the assumption $B^a = 0$.  Another possibility would 
be   to consider left and right handed tetrad fields ($A^a = - B^a$ and  
$A^a = B^a$ in (136)) and somehow to exclude one of both polarizations. 

We see that, in this way, one can write down a gravitational theory based on 
$GL(R^4)$ only, without gauging the translational group.

\section{Conclusions}

We have constructed the Higgs sector of gauge theories of gravity 
using a minimum of additional structures, and presented the 
symmetry breaking from the original symmetry group down to the 
Lorentz stability subgroup, following an approach suggested by 
the nonlinear realization treatment of those groups. The results 
of Stelle and West \cite{11} have been reproduced and the analysis has 
been extended to the Poincar\'e group, 
the affine group and the group $SL(R^5)$. 

In the case of the Poincar\'e group, the following conclusions have 
been drawn: 

1) No Poincar\'e  Lagrangian that leads to a nontrivial 
Einstein equation can be written down without the help of an additional 
Poincar\'e vector $y^a$. 

2) In order to assign a physical meaning to $y^a$, it has to be interpreted 
as 
Higgs field that triggers the symmetry breaking of the translational 
gauge freedom. 

3) No Poincar\'e invariant Higgs potential can be written down. 
This is also not necessary, since the mere presence of a Poincar\'e vector 
automatically breaks the symmetry if some groundstate value $y^a(0)$ is 
assigned to it. 

4) The most general kinetic term for $y^a$ takes the form of a cosmological 
constant. The \textit{natural}  sign of the kinetic term favors 
the choice of a  positive cosmological constant. 
A negative or zero value, however, is not  excluded on theoretical grounds. 

The following conclusions result from the analysis of the affine 
symmetry group: 

1) As in point 1 and 2 above, the general linear metric $g_{\alpha\beta}$, as 
well as the affine vector $y^{\alpha}$ are necessary to write down  
a meaningful Lagrangian for the gravitational fields and they should be 
treated as Higgs fields. 

2) The translational part of the affine group is treated in exactly the 
same way as in the case of the Poincar\'e group. Consequently, 
conclusions 3 and 4 above are valid also in this case. 

3) The symmetry breakdown of the special linear group 
$SL$ down to the Lorentz group is 
easily triggered by a Higgs sector that assigns a negative value 
to the metric determinant in the groundstate. This fixes  
the metric signature to the Lorentzian one 
and therefore fixes the stability subgroup to $O(3,1)$. The same 
signature is passed on to the physical spacetime metric. Matter 
fields that couple directly to the symmetric part of the $GL$ connection 
(i.e., possessing hypermomentum) may provoke  a signature change. 

4) In order to generalize the above procedure to the general linear 
group $GL$, the introduction of an additional \textit{dilaton} field, 
a scalar density of weight $+1$, cannot be avoided. The reason for 
this is traced back to the fact that, in metric affine theory, one of 
the equations obtained from the variation with respect to the tetrad 
and the metric respectively, is always redundant. This has its origin 
in the pure gauge nature of the tetrad field, which can be gauged into the 
non dynamical form $\delta^{\alpha}_i$ by a $GL$ transformation. As a 
result, we cannot get an independent equation for $g_{\alpha\beta}$
that would allow us to fix the groundstate of the theory. This problem 
can only be resolved by the introduction of an additional matter field, 
the dilaton $\phi$, which is 
conveniently chosen to transform as scalar density, 
and provides us with an additional, independent equation. 

5) The introduction of a Higgs sector for the dilaton field, 
initially intended to 
break down the $GL$ group to $SL$, in an approach similar to that of 
 Hehl et al. \cite{20}, reveals itself as not enough to assign 
a groundstate value to $\phi$. Only under the assumption that $\phi$ 
is nonzero, which implicitly means that we break the symmetry down 
to $SL$, the equations determine the sign of the metric determinant, 
and thus, lead to the final, Lorentz invariant groundstate metric. 
In other words, the dilaton field $\phi$ allows us to use the same 
procedure, as described under point 3, in the framework of the general linear 
group, without however assigning 
a nonzero value to $\phi$, which has to be assumed ad hoc.

6) We were able to present a solution to this problem, by introducing, 
in addition to the dilaton, a scalar field $\psi$. In this way, 
we get a complete symmetry breaking from $GL$ down to $O(3,1)$, without 
any further assumptions. In the absence of matter fields with 
hypermomentum, the signature will remain Lorentzian, and moreover, 
the Higgs sector will not contribute to the remaining gravitational 
equations, which means that the theory essentially reduces to Poincar\'e 
gauge theory.

7) A similar approach cannot be used to construct a theory with a 
residual $O(4)$ or $O(2,2)$ symmetry, since the knowledge of the 
metric determinant in those cases  will not uniquely specify the 
stability subgroup.  The Higgs sector of such theories would have 
to be constructed directly in terms of the signature (as a $GL$ invariant), 
which leads to highly nontrivial expressions. Thus, the Lorentz 
group, as opposed to the other rotation groups,  seems to be 
favored by nature. 

As to the group $SL(R^5)$, we concluded the following: 

1) The $SL(R^5)$ tensor $g_{AB}$ and vector $y^A$ are enough to 
trigger the breakdown of the symmetry group down to the Lorentz group. 
No additional structure is needed. 

2) Comparing with the affine case, the dilaton field appears now as 
the $g_{55}$ component of the metric tensor and the scalar field can 
be found in the quantity $g_{AB}y^A y^B$. A natural origin of those 
fields, introduced ad hoc in the affine theory, is therefore provided. 
Roughly, to the  affine Higgs 
fields $g_{\alpha\beta}, y^{\alpha}, \phi, \psi$ correspond the 
$SL(R^5)$ Higgs fields $g_{\alpha\beta}, y^{\alpha}, g_{55}, y^5$, 
whereas the remaining fields, namely  $g_{\alpha 5}$, 
are responsible for the breakdown of the additional, pseudo-inversional 
symmetry.

As a final conclusion, we see that the affine group as symmetry group 
of the gravitational interaction presents some interesting features 
that are not present in other gravitational gauge theories. These 
are its direct relation to general relativity and to metric theories 
in general, and the possibility of explaining dynamically the signature 
of physical spacetime. However, the symmetry breaking mechanism, 
in the form presented in this article, takes a more natural form 
in the framework of the theory based on the special linear group $SL(R^5)$. 
Therefore, it seems promising to take a closer look at that theory 
in future work.  

An alternative to the metric based models has been presented in 
the last section. It has been shown that a $GL(R^4)$ covariant 
tensor, which turns out to be related to the Dirac matrix $\gamma_5$,  
can be used to break the general linear group down to the group 
$SL(2,C)$, which is isomorphic to the Lorentz group. This 
model, although still incomplete in many aspects, seems especially 
promising as far as the incorporation of spinor fields is concerned. 
A general linear covariant generalization of the Dirac equation can 
easily be written down using only finite dimensional vector 
representations.

\end{document}